\documentclass[aip,jcp,reprint,twocolumn,superscriptaddress]{revtex4-2}

\usepackage{amsmath,amsthm,amsfonts,amssymb,times,bbm,graphicx,color,epsfig,bm}
\usepackage{hyperref}

\DeclareMathOperator{\tr}{tr}

\begin{document}

\title{1-matrix functional for long-range interaction energy of two hydrogen atoms}

\author{Jerzy Cioslowski}
\email{jerzy.cioslowski@usz.edu.pl}
\affiliation{Institute of Physics, University of Szczecin, Wielkopolska 15, 70-451 Szczecin, Poland}

\author{Christian Schilling}
\email{c.schilling@lmu.de}
\affiliation{Department of Physics, Arnold Sommerfeld Center for Theoretical Physics, Ludwig-Maximilians-Universit\"at M\"unchen, Theresienstrasse 37, 80333 M\" unchen, Germany}
\affiliation{Munich Center for Quantum Science and Technology (MCQST), Schellingstrasse 4, 80799 M\"unchen, Germany}

\author{Rolf Schilling}
\email{rschill@uni-mainz.de}
\affiliation{Institute of Physics, Johannes Gutenberg University, Staudinger Weg 9, 55099 Mainz, Germany}

\date{\today}

\begin{abstract}
The leading terms in the large-$R$ asymptotics of the functional of the one-electron reduced density matrix for the ground-state energy of the H$_2$ molecule with the internuclear separation $R$ is derived thanks to the solution of the phase dilemma at the $R \to \infty$ limit.  At this limit, the respective natural orbitals (NOs) are given by symmetric and antisymmetric combinations of “half-space” orbitals with the corresponding natural amplitudes of the same amplitudes but opposite signs.  Minimization of the resulting explicit functional yields the large-$R$ asymptotics for the occupation numbers of the weakly occupied NOs and the $C_6$ dispersion coefficient. The highly accurate approximates for the radial components of the $p$-type “half-space” orbitals and the corresponding occupation numbers (that decay like $R^{-6}$), which are available for the first time thanks to the development of the present formalism, have some unexpected properties.
\end{abstract}


\maketitle
\section{Introduction}

Dispersion (also called London or van der Waals \cite{1}) interactions are as ubiquitous in nature as their stronger counterparts commonly known as chemical bonds. However, whereas chemical bonding stems from a combination of various physical phenomena (such as electron sharing, electrostatic interactions, and polarization), a uniform mechanism, namely the long-range electron correlation, underlies all dispersion interactions.  Within the nonrelativistic clamped-nuclei approximation, this correlation gives rise to the energy lowering that for large values of the intersystem separation $R=|\vec R|$  is dominated by the
$- C_6 \, R^{-6}$ term.  Curiously, this asymptotics of the dispersion interaction energy is a universal property of Coulombic species that is independent of both the charges and the spin statistics of the particles comprising the systems $\mathcal{A}$ and $\mathcal{B}$ in question \cite{2}.
For electronic systems, application of the second-order perturbation theory produces the exact expression \cite{4,5}
\begin{eqnarray} \label{a1}
C_6 &=&  \frac{1}{2 \,\pi} \, \int_0^{\infty}
  \tr \Big[ \bm{\alpha}_\mathcal{A}(i \, \omega)  \, \Big( \frac{3 \, \vec R \otimes \vec R}{R^2} - \bm{1} \Big)
 \nonumber\\
&\times& \bm{\alpha}_\mathcal{B}(i \, \omega)
\, \Big( \frac{3 \, \vec R \otimes \vec R}{R^2} - \bm{1} \Big) \Big] \; d\omega
  \ ,
\end{eqnarray}
where $\bm{\alpha}_\mathcal{A}(i \, \omega)$ and $\bm{\alpha}_\mathcal{B}(i \, \omega)$ are the dipole dynamic polarizabilities  of $\mathcal{A}$ and $\mathcal{B}$ evaluated at imaginary frequencies (note that here and in the following the atomic units are employed), $\bm{1}$ is the unit matrix, and $\otimes$ denotes the direct product.  This expression, which for spherically symmetric $\mathcal{A}$ and $\mathcal{B}$ [i.e. those with
$\bm{\alpha}_\mathcal{X}(i \, \omega) = \alpha_\mathcal{X}(i \, \omega) \, \bm{1}$, where $\mathcal{X}$ stands for either $\mathcal{A}$ or $\mathcal{B}$] assumes the more familiar form
\begin{eqnarray}\label{a2}
C_6 =  \frac{3}{\pi} \, \int_0^{\infty}  \alpha_\mathcal{A}(i \, \omega)  \, \alpha_\mathcal{B}(i \, \omega) \; d\omega  \quad  ,
\end{eqnarray}
is not readily amenable to accurate numerical evaluation.  Consequently, a plethora of approximate schemes have been proposed, among which the well-known London approximation [which amounts to modeling $\alpha_\mathcal{X}(i \, \omega)$ with
$\frac{\alpha_\mathcal{X}(0)}{1+(\omega/I_\mathcal{X})^2}$, where $I_\mathcal{X}$ is the ionization energy of $\mathcal{X}$] \cite{6,7} and the recent variational formalism \cite{8} are worth mentioning.

The simplest case of both $\mathcal{A}$ and $\mathcal{B}$ being hydrogen atoms in their ground states has attracted considerable attention among physicists and quantum chemists alike.  In particular, numerous approaches to exact calculation of the $C_6$ coefficient have been published, including the
\textit{tour de force} of Eisenchitz and London (who have arrived at the estimate of $C_6 \approx 6.47$) \cite{9}, the variational method of Slater and Kirkwood \cite{10} (who have found $C_6 \approx 6.14$, later corrected to $C_6 \approx 6.23$ \cite{11}) that yields a two-dimensional partial differential equation approximately solvable by means of expansion into orthogonal polynomials \cite{11,12}, another variational method of Pauling and Beach (who have obtained $C_6 \approx 6.499 03$) \cite{13} that has been elaborated further by Chan and Dalgarno \cite{14}, Hirschfelder and L\"owdin \cite{15}, and Bell \cite{16}, evaluation of the dipole dynamic polarizabilities that enter Eq.~(\ref{a2}) followed by the pertinent integration \cite{17,18}, and the aforementioned
variational formalism \cite{8}.

The recently renewed interest in the one-electron reduced density matrix functional theory (1-RDMFT) has prompted research yielding mixed results.  On one hand, rigorous extensions of this approach, in which the ground-state energy of a system comprising charged fermions interacting with an external potential is given by a functional of the respective one-electron reduced density matrix (also known as the 1-matrix) \cite{19,20,21,22,23}, have been formulated, extending its applicability to excited states \cite{24} and bosonic systems (in both the ground \cite{25} and excited \cite{26} states).  On the other, several shortcomings of the existing approximate functionals \cite{27,28} as well as certain fundamental problems such as the distinction between the pure and ensemble density 1-matrix functional \cite{29} and the so-called ''phase dilemma'' \cite{30} have been exposed.

The phase dilemma arises from the lack of the one-to-one correspondence between the nonnegative-valued occupation numbers $\{ \nu_{\mathfrak{n}} \}$ of the natural (spin)orbitals (NOs) $\{ \psi_{\mathfrak{n}}(\vec r) \}$ and the potentially complex-valued linear combination coefficients in the expansion of the electronic wavefunction in terms of the Slater determinants composed from those orbitals.  As the occupation numbers are bilinear in these coefficients, the information about the sign (or, in general, phase) of these coefficients is lost.  Although the issue of the phase dilemma obviously does not contradict the existence of the 1-matrix functional for the ground-state energy, it introduces the minimization of the energy with respect to the phases in its definition, which is prohibitively expensive in terms of computational effort.  This problem is already manifest in the simplest correlated species, i.e. two-electron systems in the singlet ground states.

Bearing these facts in mind, we investigate in this paper a description of the dispersion interactions between two hydrogen atoms that is based upon 1-RDMFT.  Its advantages are several.  First of all, it involves a straightforward derivation of the 1-matrix functional for the ground-state energy whose minimization directly leads to the pertinent NOs, which are known to furnish very compact approximate electronic wavefunctions capable of reproducing the $C_6$ coefficient with reasonable accuracy \cite{15}.  Second, it provides an open-ended method for computing $C_6$ with arbitrary number of significant digits.  Third, it demonstrates how the phase dilemma can be in some cases circumvented with simple reasoning.  Fourth, it reveals the transition from the power-law decay of the occupation numbers of the NOs \cite{31,32,33} to its subexponential counterpart triggered by the predominance of the long-range correlation (entanglement) effects.

\section{Theory}

The square-normalized spatial component $\Psi(\vec r_1,\vec r_2)$ of the electronic wavefunction describing the singlet ground state of a two-electron system is given by \cite{34}
\begin{eqnarray}\label{a3}
\Psi(\vec r_1,\vec r_2) = \sum_{\mathfrak{n}} \, \lambda_{\mathfrak{n}} \, \psi_{\mathfrak{n}}(\vec r_1) \, \psi_{\mathfrak{n}}(\vec r_2)  \quad  ,
\end{eqnarray}
where the natural amplitudes (NAs) $\{ \lambda_{\mathfrak{n}} \}$ are related to the occupation numbers (per spin) of the NOs via the condition
$\forall_\mathfrak{n} \, |\lambda_{\mathfrak{n}}|^2 = \nu_{\mathfrak{n}}$  In the absence of magnetic field, both the NOs and NAs are real-valued, which implies $\forall_\mathfrak{n} \, \lambda_{\mathfrak{n}} = \sigma_{\mathfrak{n}} \, \sqrt{\nu_{\mathfrak{n}}}$, where the phases
$\{ \sigma_{\mathfrak{n}} \}$ assume the values $\pm 1$.  By convention, the NO with the largest occupation number is assigned the $+1$ phase.  The index $\mathfrak{n}$ can be either the ordinal number itself (i.e. $\mathfrak{n} =1, \, 2\, \dots$) or a combination of the ordinal number
$n = 1, \, 2\, \dots$ and some other quantum numbers.  In either case, the NOs are ordered in such a way that their occupation numbers do not increase with $\mathfrak{n}$ or $n$.   The normalization $\int |\Psi(\vec r_1,\vec r_2)|^2 \, d\vec r_1 \, d\vec r_2 =1$ implies $\sum_{\mathfrak{n}} \, \lambda_{\mathfrak{n}}^2 = \sum_{\mathfrak{n}} \, \nu_{\mathfrak{n}} = 1$.

The expectation value of the Hamiltonian $\hat H = \hat h_1 + \hat h_2 + \hat w_{12}$, where $w_{12}=|\vec r_1 - \vec r_2|^{-1}$ and the core hamiltonian
$\hat h_i = -\frac{1}{2} \, \hat \nabla_i^2 + \hat V(\vec r_i)$ involves the external potential $V(\vec r)$, reads


\begin{widetext}
\begin{eqnarray}\label{a4}
E[\Psi(\vec{r}_1,\vec{r}_2)]  &=&  \langle \Psi(\vec{r}_1,\vec{r}_2) | \hat H | \Psi(\vec{r}_1,\vec{r}_2) \rangle \\
&=& 2  \sum_{\mathfrak{n}}\, \lambda_{\mathfrak{n}}^2 \, \langle \psi_{\mathfrak{n}}(\vec{r}_1) | \hat h_1 | \psi_{\mathfrak{n}}(\vec{r}_1) \rangle + \sum_{\mathfrak{n}} \sum_{\mathfrak{n'}} \, \lambda_{\mathfrak{n}} \, \lambda_{\mathfrak{n'}}
\langle \psi_{\mathfrak{n}}(\vec{r}_1) \, \psi_{\mathfrak{n}}(\vec{r}_2) |\hat w_{12}| \psi_{\mathfrak{n'}}(\vec{r}_1)  \psi_{\mathfrak{n'}}(\vec{r}_2) \rangle \nonumber
\end{eqnarray}
which translates into the 1-matrix functional (in terms of $\{ \nu_{\mathfrak{n}} \}$ and $\{ \psi_{\mathfrak{n}}(\vec r) \}$)
\begin{equation}\label{a5}
 \mathcal{E}[\{ \nu_{\mathfrak{n}} \},\{ \psi_{\mathfrak{n}}(\vec r) \}]  =
2 \sum_{\mathfrak{n}}\, \nu_{\mathfrak{n}} \, \langle \psi_{\mathfrak{n}}(\vec r_1) | \hat h_1 | \psi_{\mathfrak{n}}(\vec r_1) \rangle
 + \min_{\{ \sigma_{\mathfrak{n}} = \pm 1 \}} \sum_{\mathfrak{n}} \sum_{\mathfrak{n'}}
\sigma_{\mathfrak{n}} \sigma_{\mathfrak{n'}}  \sqrt{\nu_{\mathfrak{n}} \, \nu_{\mathfrak{n'}}}
 \langle \psi_{\mathfrak{n}}(\vec r_1) \, \psi_{\mathfrak{n}}(\vec r_2) |\hat w_{12}| \psi_{\mathfrak{n'}}(\vec r_1) \, \psi_{\mathfrak{n'}}(\vec r_2) \rangle\,.
\end{equation}
\end{widetext}

The sign pattern of the phases $\{ \sigma_{\mathfrak{n}} \}$ that yields the global minimum in the above expression is unknown in general.  The normal sign pattern of only one positive phase \cite{35}, whose universality has been wrongly conjectured \cite{36}, is broken in several systems such as the two-electron harmonium atom with sufficiently small confinement strength $\omega $ \cite{37,38}, the members of the helium isoelectronic series with sufficiently small nuclear charge \cite{39}, and the H$_2$ molecule with large internuclear distance \cite{15,40,41,42,43}.  In rare cases, however, this sign pattern can be rigorously determined.  Thus far, these cases have comprised the harmonium atom at $\omega \ge \frac{1}{2}$ (the normal sign pattern) \cite{35} and at the limit of $\omega \to 0$ (an alternating signs pattern) \cite{37}.  In the next subsection of this paper, a new general class of systems is identified, for which certain (asymptotically valid) inferences about the sign pattern are possible.

\subsection{ The 1-matrix functional for the energy of the singlet ground state of a two-electron system with a plane of symmetry}

Consider a system in which the external potential $V(\vec r)$ satisfies the identity $V(\vec r)=V(\vec r^{\;\bullet})$ [here, and in the following, the convenient notation $\vec r^{\;\bullet} = \vec r -2 \, (\vec e_z \cdot \vec r) \, \vec e_z$, where $\vec e_z =(0,0,1)$, is employed for the vector related to $\vec r$ by the reflection with respect to the $z=0$ plane of symmetry that divides the Cartesian space into two subspaces].  Thanks to the condition
$\forall_{\mathfrak{n}} \, \psi_{\mathfrak{n}}(\vec r^{\;\bullet}) =  \eta_{\mathfrak{n}} \, \psi_{\mathfrak{n}}(\vec r)$, the NOs of such a system acquire the parity numbers $\{ \eta_{\mathfrak{n}} \}$ equal to $\pm 1$.

The probability of two electron simultaneously positioned within either of the aforementioned subspaces is given by
\begin{widetext}
\begin{equation} \label{a6}
\mathfrak{P} = \int \hspace{-4pt} \int |\Psi(\vec r_1,\vec r_2)|^2 \,
[\theta(-\vec e_z \cdot \vec r_1) \, \theta(-\vec e_z \cdot \vec r_2) + \theta(\vec e_z \cdot \vec r_1) \, \theta(\vec e_z \cdot \vec r_2)] \; d \vec r_1 \; d \vec r_2\
\end{equation}
\end{widetext}
where $\theta(x)$ is the Heaviside step function, equal to 1 for $x > 0$ and to 0 otherwise.  Combining Eqs.~(\ref{a3}) and (\ref{a6}) produces
\begin{eqnarray}\label{a7}
\mathfrak{P} = \sum_{\mathfrak{n}} \sum_{\mathfrak{n'}} \, \lambda_{\mathfrak{n}}  \, \lambda_{\mathfrak{n'}} \,
\big[ (S_{\mathfrak{n} \mathfrak{n'}}^{-})^2 + (S_{\mathfrak{n} \mathfrak{n'}}^{+})^2 \big]  \quad  ,
\end{eqnarray}
where the subspace overlaps $\{ S_{\mathfrak{n} \mathfrak{n'}}^{\pm} \}
= \{ \langle \psi_{\mathfrak{n}}(\vec r) | \theta(\pm \, \vec e_z \cdot \vec r) | \psi_{\mathfrak{n'}}(\vec r) \rangle \}$ have the property
\begin{eqnarray}\label{a8}
\mathop{{\mbox{\Large $\mathsurround0pt\forall$}}}_{\mathfrak{n} \, \mathfrak{n'}} \; S_{\mathfrak{n} \mathfrak{n'}}^{-} =
\delta_{\mathfrak{n} \mathfrak{n'}} - S_{\mathfrak{n} \mathfrak{n'}}^{+}
= \eta_{\mathfrak{n}} \, \eta_{\mathfrak{n'}} \, S_{\mathfrak{n} \mathfrak{n'}}^{+}   \quad  ,
\end{eqnarray}
from which it follows that
\begin{eqnarray}\label{a9}
 (S_{\mathfrak{n} \mathfrak{n'}}^{-})^2 + (S_{\mathfrak{n} \mathfrak{n'}}^{+})^2
&=& \frac{1}{2} \; \big[ \delta_{\mathfrak{n} \mathfrak{n'}} - 2 \, S_{\mathfrak{n} \mathfrak{n'}}^{-} \, S_{\mathfrak{n} \mathfrak{n'}}^{+} \nonumber \\
&+& (\delta_{\mathfrak{n} \mathfrak{n'}}-S_{\mathfrak{n} \mathfrak{n'}}^{+})^2 + (S_{\mathfrak{n} \mathfrak{n'}}^{+})^2 \big] \nonumber \\
&=& \frac{1}{2} \; \delta_{\mathfrak{n} \mathfrak{n'}} + (1-\eta_{\mathfrak{n}} \, \eta_{\mathfrak{n'}}) \,
(S_{\mathfrak{n} \mathfrak{n'}}^{+})^2 \ ,
\end{eqnarray}
as $\mathop{{\mbox{\large $\mathsurround0pt\forall$}}}_{\mathfrak{n}} \, S_{\mathfrak{n} \mathfrak{n}}^{-} = S_{\mathfrak{n} \mathfrak{n}}^{+} = \frac{1}{2}$.  Consequently,
\begin{eqnarray}\label{a10}
\mathfrak{P} = \frac{1}{2} + \sum_{\mathfrak{n}} \sum_{\mathfrak{n'}} \, \lambda_{\mathfrak{n}}  \, \lambda_{\mathfrak{n'}} \,
(1-\eta_{\mathfrak{n}} \, \eta_{\mathfrak{n'}}) \, (S_{\mathfrak{n} \mathfrak{n'}}^{+})^2   \quad  ,
\end{eqnarray}
i.e. only the NO pairs comprising orbitals with different parities contribute to $\mathfrak{P}-\frac{1}{2}$.

It is convenient at this point to switch to another indexing scheme, in which the ordinal number $\mathfrak{n}$ becomes the combination of the new ordinal number $n$ and the parity $\eta_n$ (written as $\bar 1$/$1$ in the subscripts).  With this scheme, Eq. (\ref{a10}) becomes simply
\begin{eqnarray}\label{a11}
 \mathfrak{P} &=& \frac{1}{2} + 4 \, \sum_{n} \sum_{n'} \, \lambda_{n 1}  \, \lambda_{n' \bar 1} \,  (S_{n 1,n' \bar 1}^{+})^2 \nonumber \\
&=&   2 \, \sum_{n} \sum_{n'} \, (\lambda_{n 1}  + \lambda_{n' \bar 1})^2 \,  (S_{n 1,n' \bar 1}^{+})^2  \nonumber \\
&+& \sum_{n} \lambda_{n 1}^2  \, \langle \xi_{n 1}(\vec r) | \theta(\vec e_z \cdot \vec r) | \xi_{n 1}(\vec r) \rangle
\nonumber \\
&+& \sum_{n'} \, \lambda_{n' \bar 1}^2 \,  \langle \xi_{n' \bar 1}(\vec r) | \theta(\vec e_z \cdot \vec r) | \xi_{n' \bar 1}(\vec r) \rangle \;  ,
\end{eqnarray}
where (note the new indexing scheme for the NOs)
\begin{eqnarray}\label{a12}
\xi_{n 1}(\vec r) =  \psi_{n 1}(\vec r) - 2 \, \sum_{n'} \, S_{n 1,n' \bar 1}^{+} \, \psi_{n' \bar 1}(\vec r)   \quad  ,
\end{eqnarray}
\begin{eqnarray}\label{a13}
\xi_{n' \bar 1}(\vec r) =  \psi_{n' \bar 1}(\vec r) - 2 \, \sum_{n} \, S_{n 1,n' \bar 1}^{+} \, \psi_{n 1}(\vec r)   \quad ,
\end{eqnarray}
and the following identities have been employed: $ \sum_{n} \lambda_{n 1}^2 + \sum_{n'} \, \lambda_{n' \bar 1}^2 =1 $,
$\mathop{{\mbox{\large $\mathsurround0pt\forall$}}}_{n n'} \, S_{n 1,n' 1}^{+} = S_{n \bar 1,n' \bar 1}^{+} = \frac{1}{2} \, \delta_{nn'}$,
$\mathop{{\mbox{\large $\mathsurround0pt\forall$}}}_{n} \, \sum_{n'} \, (S_{n 1,n' \bar 1}^{+})^2 = \frac{1}{4} - \frac{1}{2} \, \langle \xi_{n 1}(\vec r) | \theta(\vec e_z \cdot \vec r) | \xi_{n 1}(\vec r) \rangle$ and
$\mathop{{\mbox{\large $\mathsurround0pt\forall$}}}_{n'} \, \sum_{n} \, (S_{n 1,n' \bar 1}^{+})^2 = \frac{1}{4} - \frac{1}{2} \, \langle \xi_{n' \bar 1}(\vec r) | \theta(\vec e_z \cdot \vec r) | \xi_{n' \bar 1}(\vec r) \rangle$.

It follows from Eq. (\ref{a11}) that $\mathfrak{P}$ is a sum of three nonnegative-valued terms.  Therefore, $\mathfrak{P}=0$ only if each of these terms vanishes individually.  For the first term, this implies that for every pair $(n, n')$ of indices, $\lambda_{n 1}  + \lambda_{n' \bar 1}$ and $S_{n 1,n' \bar 1}^{+}$ cannot simultaneously assume nonzero values.  For the other two terms, it imposes the conditions
$\mathop{{\mbox{\large $\mathsurround0pt\forall$}}}_{n} \, \sum_{n'} \, (S_{n 1,n' \bar 1}^{+})^2 = \frac{1}{4}$ and
$\mathop{{\mbox{\large $\mathsurround0pt\forall$}}}_{n'} \, \sum_{n} \, (S_{n 1,n' \bar 1}^{+})^2 = \frac{1}{4}$ that stem from the last two of the above identities and the nonnegative valuedness of $\langle \xi_{n 1}(\vec r) | \theta(\vec e_z \cdot \vec r) | \xi_{n 1}(\vec r) \rangle$ and
$\langle \xi_{n' \bar 1}(\vec r) | \theta(\vec e_z \cdot \vec r) | \xi_{n' \bar 1}(\vec r) \rangle$.  Consequently, for each $n$ there has to be at least one nonzero-valued $S_{n 1,n' \bar 1}^{+}$ and thus at least one $n'$ for which $\lambda_{n 1}  + \lambda_{n' \bar 1}=0$.  In fact, there has to be exactly one such $n'$ for each $n$, as having $\lambda_{n 1}  + \lambda_{n'_1 \bar 1}=0$ and $\lambda_{n 1}  + \lambda_{n'_2 \bar 1}=0$ for two different $n'_1$ and $n'_2$ would imply $\lambda_{n'_1 \bar 1}=\lambda_{n'_2 \bar 1}$, i.e. introduce spurious degeneracies among the NOs.  Since the occupation numbers of the NOs are ordered nonascendingly, the pairing of the NAs occurs at $n'=n$.  This means that
$\mathop{{\mbox{\large $\mathsurround0pt\forall$}}}_{n n'} \, S_{n 1,n' \bar 1}^{+} = S_{n 1,n \bar 1}^{+} \, \delta_{nn'}$ and
$\mathop{{\mbox{\large $\mathsurround0pt\forall$}}}_{n} \, (S_{n 1,n \bar 1}^{+})^2 = \frac{1}{4}$ (which, without any loss of generality, can be replaced with $\mathop{{\mbox{\large $\mathsurround0pt\forall$}}}_{n} \, S_{n 1,n \bar 1}^{+} = \frac{1}{2}$).  Consequently, one infers from Eq.~(\ref{a12}) that
$\mathop{{\mbox{\large $\mathsurround0pt\forall$}}}_{n} \, \xi_{n 1}(\vec r) =  \psi_{n 1}(\vec r) - \psi_{n \bar 1}(\vec r)$, which in conjunction with the vanishing of $\langle \xi_{n 1}(\vec r) | \theta(\vec e_z \cdot \vec r) | \xi_{n 1}(\vec r) \rangle$ produces
$\mathop{{\mbox{\large $\mathsurround0pt\forall$}}}_{n} \, \mathop{{\mbox{\large $\mathsurround0pt\forall$}}}_{\vec e_z \cdot \vec r \ge 0} \,
\psi_{n 1}(\vec r) - \psi_{n \bar 1}(\vec r) = 0$.

In summary, the vanishing of $\mathfrak{P}$ pertaining to the wavefunction (\ref{a3}) implies perfect pairing of the NOs with opposite parities and the corresponding NAs, namely
\begin{eqnarray}\label{a14}
\mathop{{\mbox{\LARGE $\mathsurround0pt\forall$}}}_{n} \; \Big( \lambda_{n 1}  + \lambda_{n \bar 1} = 0  \;  \land \;
\theta(\vec e_z \cdot \vec r) \, [\psi_{n 1}(\vec r) - \psi_{n \bar 1}(\vec r)] = 0\,. \nonumber\\
\end{eqnarray}
The first of these pairings imposes the conditions $\forall_{n} \, \nu_{n 1} =\nu_{n \bar 1}$ and $\forall_{n} \, \sigma_{n 1} = - \sigma_{n \bar 1}$ upon the occupation numbers and the phases, respectively.  A practical way of interpreting the second of these pairings involves employing the representation
\begin{widetext}
\begin{equation} \label{a15}
\mathop{{\mbox{\LARGE $\mathsurround0pt\forall$}}}_{n} \; \Big( \psi_{n 1}(\vec r) = \frac{1}{\sqrt{2}} \; [\phi_n(\vec r)+\phi_n(\vec r^{\;\bullet})]
\;  \land \;  \psi_{n \bar 1}(\vec r) = \frac{1}{\sqrt{2}} \; [\phi_n(\vec r)-\phi_n(\vec r^{\;\bullet})] \, \Big)  \  ,
\end{equation}
\end{widetext}
where the orthonormal ''half-space orbitals'' $\{ \phi_n(\vec r) \}$ have the $z>0$ subspace as their support, i.e. they conform to
$\mathop{{\mbox{\large $\mathsurround0pt\forall$}}}_{n} \, \theta(-\vec e_z \cdot \vec r) \, \phi_{n}(\vec r) = 0$.  This condition implies the ZDO (zero-differential overlap) property of $\mathop{{\mbox{\large $\mathsurround0pt\forall$}}}_{n} \, \phi_n(\vec r) \, \phi_n(\vec r^{\;\bullet}) = 0$.

For a small positive-valued $\mathfrak{P}$, the equality signs in Eqs. (\ref{a14}) and (\ref{a15}), as well as in the two conditions for $\{ \phi_n(\vec r) \}$ that follow, have to be replaced with $\approx$ understood as ''$A \approx B$ means $A-B$ is of the order of some power of $\mathfrak{P}$'' (e.g.
$\mathop{{\mbox{\large $\mathsurround0pt\forall$}}}_{n} \, |\lambda_{n 1}  + \lambda_{n \bar 1}| \le \sqrt{2 \, \mathfrak{P}}$, etc.).  Thus, applying the ZDO property while combining Eqs. (\ref{a5}), (\ref{a14}), and (\ref{a15}) produces

\begin{eqnarray}\label{a16}
\mathcal{E}[\{ \nu_{\mathfrak{n}} \},\{ \psi_{\mathfrak{n}}(\vec r) \}]  &\approx&
\mathcal{\tilde E}[\{ \nu_{\mathfrak{n}} \},\{ \phi_{\mathfrak{n}}(\vec r) \}] \\
&=& 4 \, \sum_{\mathfrak{n}}\, \nu_{\mathfrak{n}} \, \langle \phi_{\mathfrak{n}}(\vec r_1) | \hat h_1 | \phi_{\mathfrak{n}}(\vec r_1) \rangle \nonumber \\
&+& 2 \, \min_{\{ \sigma_{\mathfrak{n}} = \pm 1 \}} \; \sum_{\mathfrak{n}} \sum_{\mathfrak{n'}} \,
\sigma_{\mathfrak{n}} \, \sigma_{\mathfrak{n'}} \, \sqrt{\nu_{\mathfrak{n}} \, \nu_{\mathfrak{n'}}}
 \nonumber\\
&\times&  \langle \phi_{\mathfrak{n}}(\vec r_1) \, \phi_{\mathfrak{n}}(\vec r_2^{\;\bullet}) |\hat w_{12}|
\phi_{\mathfrak{n'}}(\vec r_1) \, \phi_{\mathfrak{n'}}(\vec r_2^{\;\bullet}) \rangle   \ , \nonumber
\end{eqnarray}
where the functional $\mathcal{\tilde E}[\{ \nu_{\mathfrak{n}} \},\{ \phi_{\mathfrak{n}}(\vec r) \}]$ yields the leading term in the
$\mathfrak{P} \to 0$ energy asymptotics in terms of the half-space orbitals $\{ \phi_{\mathfrak{n}}(\vec r) \}$ and their amplitudes $\{ \nu_{\mathfrak{n}} \}$ [note the updated sum rule $\sum_{\mathfrak{n}} \, \nu_{\mathfrak{n}} = \frac{1}{2}$ and the replacement of $n$ with $\mathfrak{n}$ as the ordinal number (different from that employed previously) in anticipation of further developments in Section II.3 of this paper].

\subsection{ The 1-matrix functional for the energy of the singlet ground state of the H$_2$ molecule with large internuclear distance}

The external potential pertaining to the H$_2$ molecule with nuclei positioned at $\pm \frac{1}{2} \, R \, \vec e_z$ is given by
$V(\vec r) = - \big( |\vec r - \frac{1}{2} \, R \, \vec e_z|^{-1} + |\vec r + \frac{1}{2} \, R \, \vec e_z|^{-1} \big)$. Since $\mathfrak{P}$ is known to decay exponentially with sufficiently large $R$ \cite{44}, the considerations of the previous subsection are relevant to this system, for which it is convenient to define
the shifted half-space orbitals $\{ \bar\phi_{\mathfrak{n}}(\vec r) \} = \{ \phi_{\mathfrak{n}}(\vec r + \frac{1}{2} \, R \, \vec e_z) \}$.  The corresponding functional reads
\begin{widetext}
\begin{eqnarray}\label{a17}
 \mathcal{\bar E}[\{ \nu_{\mathfrak{n}} \},\{ \bar\phi_{\mathfrak{n}}(\vec r) \}] &=&
4 \, \sum_{\mathfrak{n}}\, \nu_{\mathfrak{n}} \, \Big\langle \bar\phi_{\mathfrak{n}}(\vec r) \Big|
 -\frac{1}{2} \, \hat \nabla^2 - \big( |\vec r|^{-1} + |\vec r + R \, \vec e_z|^{-1} \big) \, \hat 1 \Big| \bar\phi_{\mathfrak{n}}(\vec r) \Big\rangle   \nonumber \\
&+&  \, 2 \, \min_{\{ \sigma_{\mathfrak{n}} = \pm 1 \}} \; \sum_{\mathfrak{n}} \sum_{\mathfrak{n'}} \,
\sigma_{\mathfrak{n}} \, \sigma_{\mathfrak{n'}} \, \sqrt{\nu_{\mathfrak{n}} \, \nu_{\mathfrak{n'}}}
  \big\langle \bar\phi_{\mathfrak{n}}(\vec r_1) \, \bar\phi_{\mathfrak{n}}(\vec r_2^{\;\bullet}) \big|
|\vec r_1-\vec r_2+ R \, \vec e_z|^{-1} \big| \bar\phi_{\mathfrak{n'}}(\vec r_1) \, \bar\phi_{\mathfrak{n'}}(\vec r_2^{\;\bullet}) \big\rangle \quad .
\end{eqnarray}
\end{widetext}
The two-electron integrals that enter Eq. (\ref{a17}) are amenable to power expansion in terms of $R^{-1}$ upon application of the multipole expansion \cite{45}
\begin{eqnarray}\label{a18}
 |\vec r_1-\vec r_2+ R \, \vec e_z|^{-1} &=& \sum_{l_1=0}^{\infty} \, \sum_{l_2=0}^{\infty} \; \sum_{m=-\min(l_1,l_2)}^{\min(l_1,l_2)} \,
\, (-1)^{l_2} \, C_{l_1,l_2,m} \;   \nonumber \\
 &\times&   \frac{r_1^{l_1} r_2^{l_2}}{R^{l_1+l_2+1}} Y_{l_1}^m(\theta_1,\varphi_1) \, Y_{l_2}^{-m}(\theta_2,\varphi_2)   , \nonumber \\
\end{eqnarray}
where
\begin{eqnarray}\label{a19}
 C_{l_1,l_2,m} &&= 4 \, \pi   \, (l_1+l_2)! \big[ (2 \, l_1+1) \, (2 \, l_2+1) \, (l_1-m)! \, \nonumber \\
 &\times& \,  \, (l_2-m)! (l_1+m)! \, (l_2+m)! \big]^{-1/2}   \quad ,
\end{eqnarray}
$Y_l^m(\theta,\varphi)$ is the spherical harmonic, $\vec r_1 = r_1 \, (\sin\theta_1 \, \cos\varphi_1, \sin\theta_1 \, \sin\varphi_1, \cos\theta_1)$,
and $\vec r_2 = r_2 \, (\sin\theta_2 \, \cos\varphi_2, \sin\theta_2 \, \sin\varphi_2, \cos\theta_2)$.  Since the terms with $l_1=0$ or $l_2=0$ cancel out with the third term in the core Hamiltonian expectation value whereas the $l_1=l_2=0$ term cancels out partially, one obtains
\begin{widetext}
\begin{eqnarray}\label{a20}
 \mathcal{\bar E}_L[\{ \nu_{\mathfrak{n}} \},\{ \bar\phi_{\mathfrak{n}}(\vec r) \}] &=&
4 \, \sum_{\mathfrak{n}}\, \nu_{\mathfrak{n}} \, \Big\langle \bar\phi_{\mathfrak{n}}(\vec r) \Big|
 -\frac{1}{2} \, \hat \nabla^2 - \big[ |\vec r|^{-1} + (2 \, R)^{-1}) \big] \, \hat 1 \Big| \bar\phi_{\mathfrak{n}}(\vec r) \Big\rangle + \nonumber \\
&+&  \, 2 \, \min_{\{ \sigma_{\mathfrak{n}} = \pm 1 \}} \;
\sum_{l_1=1}^{L-2} \, \sum_{l_2=1}^{L-l_1-1} \; \sum_{m=-\min(l_1,l_2)}^{\min(l_1,l_2)} \, C_{l_1,l_2,m} \, R^{-(l_1+l_2+1)} \nonumber\\
&\times& \sum_{\mathfrak{n}} \sum_{\mathfrak{n'}} \,
\sigma_{\mathfrak{n}} \, \sigma_{\mathfrak{n'}} \, \sqrt{\nu_{\mathfrak{n}} \, \nu_{\mathfrak{n'}}}
  \,
\langle \bar\phi_{\mathfrak{n}}(\vec r_1)|\hat r_1^{l_1} \, \hat Y_{l_1}^m(\theta_1,\varphi_1)| \bar\phi_{\mathfrak{n'}}(\vec r_1) \rangle \,
\langle \bar\phi_{\mathfrak{n}}(\vec r_2)|\hat r_2^{l_2} \, \hat Y_{l_2}^m(\theta_2,\varphi_2)| \bar\phi_{\mathfrak{n'}}(\vec r_2) \rangle^{*} \quad
\end{eqnarray}
\end{widetext}
for the functional $\mathcal{\bar E}_L[\{ \nu_{\mathfrak{n}} \},\{ \bar\phi_{\mathfrak{n}}(\vec r) \}]$ that yields the energy up to the term proportional to $R^{-L}$.  The large-$R$ asymptotics of $\mathcal{\bar E}[\{ \nu_{\mathfrak{n}} \},\{ \bar\phi_{\mathfrak{n}}(\vec r) \}]$ has the leading terms
\begin{eqnarray}\label{a21}
\mathcal{\bar E}_3[\{ \nu_{\mathfrak{n}} \},\{ \bar\phi_{\mathfrak{n}}(\vec r) \}] &=& 4 \, \sum_{\mathfrak{n}}\, \nu_{\mathfrak{n}} \, h_{\mathfrak{n}\mathfrak{n}} + 2 \, R^{-3} \times \\
&\times&  \, \min_{\{ \sigma_{\mathfrak{n}} = \pm 1 \}} \; \sum_{\mathfrak{n}} \sum_{\mathfrak{n'}} \,
\sigma_{\mathfrak{n}} \, \sigma_{\mathfrak{n'}} \, \sqrt{\nu_{\mathfrak{n}} \, \nu_{\mathfrak{n'}}} \; I_{\mathfrak{n}\mathfrak{n'}} \,, \nonumber
\end{eqnarray}
where  [note that $\vec r =(x,y,z)$]
$h_{\mathfrak{n}\mathfrak{n}} = \big\langle \bar\phi_{\mathfrak{n}}(\vec r) \big| - \frac{1}{2} \, \hat \nabla^2 - (|\vec r|^{-1} + \frac{1}{2} \, R^{-1})  \, \hat 1 \big| \bar\phi_{\mathfrak{n}}(\vec r) \big\rangle$ and
$I_{\mathfrak{n}\mathfrak{n'}} = \big( \langle \bar\phi_{\mathfrak{n}}(\vec r)|\hat x| \bar\phi_{\mathfrak{n'}}(\vec r) \rangle \big)^2
+ \big( \langle \bar\phi_{\mathfrak{n}}(\vec r)|\hat y| \bar\phi_{\mathfrak{n'}}(\vec r) \rangle \big)^2
+ 2 \, \big( \langle \bar\phi_{\mathfrak{n}}(\vec r)|\hat z| \bar\phi_{\mathfrak{n'}}(\vec r) \rangle \big)^2$.

\subsection{Resolution of the  phase dilemma for $\mathcal{\bar E}_3[\{ \nu_{\mathfrak{n}} \},\{ \bar\phi_{\mathfrak{n}}(\vec r) \}]$ }

Although the phase dilemma concerning the functional $\mathcal{\bar E}_3[\{ \nu_{\mathfrak{n}} \},\{ \bar\phi_{\mathfrak{n}}(\vec r) \}]$ with the nonnegative-valued terms $\{ I_{\mathfrak{n}\mathfrak{n'}} \}$  appears at the first glance to be analogous to that of the functional
$\mathcal{E}[\{ \nu_{\mathfrak{n}} \},\{ \psi_{\mathfrak{n}}(\vec r) \}] $ with the nonnegative-valued integrals $\{ \langle \psi_{\mathfrak{n}}(\vec r_1) \, \psi_{\mathfrak{n}}(\vec r_2) |\hat w_{12}| \psi_{\mathfrak{n'}}(\vec r_1) \, \psi_{\mathfrak{n'}}(\vec r_2) \rangle \}$, it is asymptotically tractable at the limit of $R \to \infty$.  This is so because for sufficiently large $R$ one can directly find the minimizers $\{ \sigma_n(\{\nu_n\},\{\bar\phi_n(\vec r)\}) \}$ for the second term in the r.h.s. of Eq. (\ref{a21}).  Another approach, in which this minimization is carried out in an indirect manner, is mathematically more convenient and thus employed here.

Let the matrix $\bm{\Lambda} \equiv \bm{\Lambda}(R)$ have the elements
$\Lambda_{\mathfrak{n}\mathfrak{n'}} = R^{-3} \, (\mathfrak{E} - 2 \,  h_{\mathfrak{n}\mathfrak{n}})^{-1} I_{\mathfrak{n}\mathfrak{n'}}$, where
$\{ h_{\mathfrak{n}\mathfrak{n}} \} \equiv \{h_{\mathfrak{n}\mathfrak{n}}(R) \}$ and
$\{ I_{\mathfrak{n}\mathfrak{n'}} \} \equiv \{I_{\mathfrak{n}\mathfrak{n'}}(R) \}$ are evaluated at the \textit{actual} shifted  half-space orbitals, i.e. those that minimize $\mathcal{\bar E}_3[\{ \nu_{\mathfrak{n}} \},\{ \bar\phi_{\mathfrak{n}}(\vec r) \}]$ for a given $R$. Finding the minimum
$\mathfrak{E} \equiv \mathfrak{E}(R)$ of  this functional is equivalent to solving the (infinite-dimensional) eigenproblem
$\bm{\Lambda} \, \bm{\lambda} = \bm{\lambda}$, where the vector $\bm{\lambda} \equiv \bm{\lambda}(R)$ has the elements
$\lambda_{\mathfrak{n}} = \sigma_{\mathfrak{n}} \, \sqrt{\nu_{\mathfrak{n}}}$.  Thus the signs of
$\{ \lambda_{ \mathfrak{n}} \} \equiv \{ \lambda_{ \mathfrak{n}}(R) \}$ yield the \textit{actual} phases
$\{ \sigma_{ \mathfrak{n}} \} \equiv \{ \sigma_{ \mathfrak{n}}(R) \}$.

At this point, it is convenient to change the numbering convention from $\mathfrak{n} =1, \, 2\, \dots$ to $\mathfrak{n} =0, \, 1\, \dots$ and split the above eigenproblem into
\begin{eqnarray}\label{a22}
\lambda_0 &=& (1-\Lambda_{00})^{-1} \, \sum_{\mathfrak{n}=1}^{\infty} \, \Lambda_{0\mathfrak{n}} \, \lambda_{\mathfrak{n}}  \nonumber\\
&=& [ (\mathfrak{E} - 2 \, h_{00}) \,  R^3 - I_{00} ]^{-1} \, \sum_{\mathfrak{n}=1}^{\infty} \, I_{0\mathfrak{n}} \, \lambda_{\mathfrak{n}}  \;
\end{eqnarray}
and $\bm{\tilde\Lambda} \, \bm{\tilde\lambda} = \bm{\tilde\lambda}$, where the matrix $\bm{\tilde\Lambda} \equiv \bm{\tilde\Lambda}(R)$ has the elements
\begin{eqnarray}\label{a23}
 \tilde\Lambda_{\mathfrak{n}\mathfrak{n'}} &=& \Lambda_{\mathfrak{n}\mathfrak{n'}}
+ (1-\Lambda_{00})^{-1} \, \Lambda_{\mathfrak{n}\mathfrak{0}} \, \Lambda_{\mathfrak{0}\mathfrak{n'}} = R^{-3} \, (\mathfrak{E} - 2 \,  h_{\mathfrak{n}\mathfrak{n}})^{-1} \times \nonumber \\
&\times&   \, \big( I_{\mathfrak{n}\mathfrak{n'}}
+ [ (\mathfrak{E} - 2 \, h_{00}) \,  R^3 - I_{00} ]^{-1} \, I_{\mathfrak{n}0} \, I_{0\mathfrak{n'}} \big)
\end{eqnarray}
and the vector $\bm{\tilde\lambda} \equiv \bm{\tilde\lambda}(R) \equiv (\lambda_1,\lambda_2,\dots)$ satisfies the normalization condition
$\lambda_0^2 + \bm{\tilde\lambda}^{T} \bm{\tilde\lambda} =\frac{1}{2}$.

The large-$R$ asymptotics of $\bm{\tilde \Lambda}(R)$ and $\mathfrak{E}(R)$ are readily deduced from those of $\{ h_{\mathfrak{n}\mathfrak{n}} \}$ and
$\{ I_{\mathfrak{n}\mathfrak{n'}} \}$.  Since the actual $\bar\phi_{0}(\vec r)$ converges at the dissociation limit to the 1s orbital of the hydrogen atom, all
$\{ I_{\mathfrak{n}\mathfrak{n'}} \}$ tend at $R \to \infty$ to finite nonnegative-valued constants whereas the differences $\{ h_{\mathfrak{n}\mathfrak{n}}-h_{00} \}$ tend to finite positive-valued constants for all $\mathfrak{n} \ne 0$  \cite{46}.  Having the $R$-independent eigenvalue of one,
$\bm{\tilde \Lambda}(R)$ cannot vanish at this limit.  Therefore, the expression $[ (\mathfrak{E} - 2 \, h_{00}) \,  R^3 - I_{00} ]^{-1}$ has to scale asymptotically like $R^3$, which implies the large-$R$ behavior $\mathfrak{E} = 2 \, h_{00} + I_{00} \,  R^{-3} - C_6 \, R^{-6}+ \dots$~ with $C_6$ that is positive-valued thanks to the inequality
$2 \, h_{00} + I_{00} \,  R^{-3} = \mathcal{\bar E}_3[\{ \frac{1}{2},0,\dots \},\{ \bar\phi_{\mathfrak{n}}(\vec r) \}] > \mathfrak{E}$ pertaining to the actual
$\{ \bar\phi_{\mathfrak{n}}(\vec r) \}$.  Consequently,
\begin{eqnarray}\label{a24}
\mathop{{\mbox{\LARGE $\mathsurround0pt\forall$}}}_{\mathfrak{n} \ne 0,\mathfrak{n'} \ne 0} \; \lim_{R \to \infty}
\, \tilde\Lambda_{\mathfrak{n}\mathfrak{n'}}
&=& - \frac{1}{2} \; C_6^{-1} \, (h_{00}- h_{\mathfrak{n}\mathfrak{n}})^{-1} \times \nonumber\\
&\times& I_{\mathfrak{n}0} \, I_{0\mathfrak{n'}} \ge 0 \quad .
\end{eqnarray}

Being given by a direct product of two vectors, $\bm{\tilde\Lambda}(\infty)$ is a rank-1 matrix whose sole nonzero eigenvalue equaling
$- \frac{1}{2} \, C_6^{-1} \, \sum_{\mathfrak{n}=1}^{\infty} \, (h_{00}- h_{\mathfrak{n}\mathfrak{n}})^{-1} \, I_{\mathfrak{n}0} \, I_{0\mathfrak{n}}$ and the corresponding eigenvector with the components $\{ A \, (h_{00}- h_{\mathfrak{n}\mathfrak{n}})^{-1} \, I_{\mathfrak{n}0} \}$ are the sought solutions of the eigenequation $\bm{\tilde\Lambda}(\infty) \, \bm{\tilde\lambda} = \bm{\tilde\lambda}$ provided
$C_6 = - \frac{1}{2} \, \sum_{\mathfrak{n}=1}^{\infty} \, (h_{00}- h_{\mathfrak{n}\mathfrak{n}})^{-1} \, I_{\mathfrak{n}0}^2$ (note all the pertinent matrix elements being evaluated at $R=\infty$).  The normalization factor $A$ is obtained by combining Eq.~(\ref{a22}) with the aforementioned condition for the norm of $\bm{\tilde\lambda}$, which yields
$A = 2^{-3/2} \, R^{-3} + \dots$ and $\lambda_0 = 2^{-1/2} + \dots$ for the leading terms in the respective $R \to \infty$ asymptotics.  Thus, at the limit of $R \to \infty$, $\lambda_0$ tends to $\frac{1}{\sqrt{2}}$ whereas, for all $\mathfrak{n} \ne 0$ that correspond to nonvanishing $I_{\mathfrak{n}0}$,  $\lambda_{\mathfrak{n}}$ are negative-valued and decay like $R^{-3}$.  In other words, $\nu_0$ tends to $\frac{1}{2}$ and the asymptotic equality
$\nu_ {\mathfrak{n}} =\bar \nu_ {\mathfrak{n}} \, R^{-6} + \dots$, where
$\bar \nu_ {\mathfrak{n}} = \frac{1}{8} \, (h_{00}- h_{\mathfrak{n}\mathfrak{n}})^{-2} \, I_{\mathfrak{n}0}^2$, holds for all $\mathfrak{n} \ne 0$.

By a simple continuity argument, the properties of $\bm{\tilde\Lambda}(\infty)$ that transpire from the above considerations, namely the nonnegative valuedness of its elements and its largest eigenvalue equaling one, are retained in $\bm{\tilde\Lambda}(R)$ pertaining to sufficiently large finite $R$.  By virtue of the Perron-Frobenius theorem \cite{47}, these properties imply the nonpositive valuedness of $\lambda_{\mathfrak{n}}$ for all $\mathfrak{n} \ne 0$, i.e. persistence of the normal pattern exhibited by the signs of the phases $\{ \sigma_{\mathfrak{n}} \}$ for $R > R_{crit}$, where
$R_{crit}$ is the solution of the transcendental equation
\begin{eqnarray}\label{a25}
\big[ \mathfrak{E}(R_{crit}) &-& 2 \, h_{00}(R_{crit}) \big] \,  R_{crit}^3 - I_{00}(R_{crit}) + \nonumber\\
&+& \min_{\mathfrak{n} \ne0,\mathfrak{n'} \ne 0} \,
\frac{I_{\mathfrak{n}0}(R_{crit}) \, I_{0\mathfrak{n'}}(R_{crit})}{I_{\mathfrak{n}\mathfrak{n'}}(R_{crit})} = 0 \quad \nonumber  \\
\end{eqnarray}
Two observations are in order here.  First of all, it is possible that the sign pattern of the phases remains normal at certain $R$ smaller than $R_{crit}$ as the Perron-Frobenius theory provides the sufficient but not necessary conditions for $\bm{\tilde\lambda}$ being unisigned (i.e. having components of the same sign).  Second, Eq.~(\ref{a25}) has a finite-valued solution unless the minimum [which excludes the index pairs $(\mathfrak{n},\mathfrak{n'})$ for which $I_{\mathfrak{n}\mathfrak{n'}}(R_{crit}) = 0$] that enters its l.h.s. equals zero.  In such a case, the possibility of which cannot be ruled out at present, the negative valuedness of $\sigma_{\mathfrak{n}}$ would not be maintained beyond certain $\mathfrak{n}_{crit}(R)$.

\subsection{Minimization of  $\mathcal{\bar E}_3[\{ \nu_{\mathfrak{n}} \},\{ \bar\phi_{\mathfrak{n}}(\vec r) \}]$ at $R \to \infty$}

In light of the discussion in the preceding subsection of this paper, the minimization of
$\mathcal{\bar E}_3[\{ \nu_{\mathfrak{n}} \},\{ \bar\phi_{\mathfrak{n}}(\vec r) \}]$ at $R \to \infty$ reduces to finding the maximum
\begin{widetext}
\begin{eqnarray}\label{a26}
C_6 = \frac{1}{2} \, \max_{\{ \bar\phi_{\mathfrak{n}} \}} \, \sum_{\mathfrak{n}=1}^{\infty}
\frac{\big( [\langle \chi(\vec r)|\hat x| \bar\phi_{\mathfrak{n}}(\vec r) \rangle]^2
             + [\langle \chi(\vec r)|\hat y| \bar\phi_{\mathfrak{n}}(\vec r) \rangle]^2
       + 2 \, [\langle \chi(\vec r)|\hat z| \bar\phi_{\mathfrak{n}}(\vec r) \rangle]^2 \big)^2}
{\big\langle \bar\phi_{\mathfrak{n}}(\vec r) \big| - \frac{1}{2} \, \hat \nabla^2 - |\vec r|^{-1} \, \hat 1 \big| \bar\phi_{\mathfrak{n}}(\vec r) \big\rangle+\frac{1}{2}} \quad  \hspace{-30pt} \nonumber \\
\end{eqnarray}
\end{widetext}
over the orthonormal set of the half-space orbitals $\{ \bar\phi_{\mathfrak{n}}(\vec r) \}$ orthogonal to the 1s orbital
$\chi(\vec r)=\pi^{-1/2} \, \exp(-|\vec r|)$ of the hydrogen atom.  This task is greatly simplified by the cylindrical symmetry of the H$_2$ molecule, which restricts the dependence of the real-valued $\{ \bar\phi_{\mathfrak{n}}(\vec r) \}$ on
$\vec r = r \, (\sin\theta \, \cos\varphi, \sin\theta \, \sin\varphi, \cos\theta)$ to that consistent with the proportionality to the angular factors of either
$\sin m \varphi$ (with $m=1,\, 2,\, \dots$) or $\cos m \varphi$ (with $m=0, \, 1,\, 2,\, \dots$).  Furthermore, due to the presence of the $\hat x$, $\hat y$, and $\hat z$ operators in the matrix element $I_{\mathfrak{n}0}$, the large-$R$ asymptotics of $\nu_{\mathfrak{n}}$ has the leading term proportional to $R^{-6}$ only when the angular factor in the corresponding $\{ \bar\phi_{\mathfrak{n}}(\vec r) \}$ equals 1, $\sin \varphi$, or $\cos \varphi$.  Finally, thanks to the elementary properties of spherical harmonics, the set of the possible half-space orbitals narrows down to the products
$\frac{1}{2} \, \sqrt{\frac{3}{\pi}} \; x \, \zeta_n(r)$, $\frac{1}{2} \, \sqrt{\frac{3}{\pi}} \; y \, \zeta_n(r)$, and
$\frac{1}{2} \, \sqrt{\frac{3}{\pi}} \; z \, \zeta_n(r)$ [where
$\langle \zeta_n(r)|\zeta_{n'}(r) \rangle = \int_{0}^{\infty} \zeta_n(r) \, \zeta_{n'}(r) \, r^4 \, dr =\delta_{nn'}$] corresponding to $\mathfrak{n}$ being replaced by the combinations $nx$, $ny$, and $nz$.  For a given ordinal number $n$ (which assumes the values of $1, \, 2, \dots$),  $\bar\phi_{nx}(\vec r)$, $\bar\phi_{ny}(\vec r)$, and $\bar\phi_{nz}(\vec r)$ yield the same denominators
$\big\langle \bar\phi_{\mathfrak{n}}(\vec r) \big| - \frac{1}{2} \, \hat \nabla^2 - |\vec r|^{-1}   \, \hat 1 \big| \bar\phi_{\mathfrak{n}}(\vec r) \big\rangle
+\frac{1}{2}= \frac{1}{2}-\frac{1}{2} \, \int_{0}^{\infty} \zeta_n(r) \, [(2 \, \zeta_n(r) + 4 \, \zeta_n'(r) + r \, \zeta_n''(r)] \, r^3 \, dr$ in the r.h.s. of Eq.~(\ref{a26}), whereas the corresponding numerators read $\frac{16}{9} \, \big[ \int_{0}^{\infty} \zeta_{n}(r) \, \exp(-r) \, r^4 \, dr \big]^4$,
$\frac{16}{9} \, \big[ \int_{0}^{\infty} \zeta_{n}(r) \, \exp(-r) \, r^4 \, dr \big]^4$, and
$\frac{64}{9} \, \big[ \int_{0}^{\infty} \zeta_{n}(r) \, \exp(-r) \, r^4 \, dr \big]^4$, respectively.  Consequently, Eq.~(\ref{a26}) is equivalent to
\begin{widetext}
\begin{eqnarray}\label{a27}
C_6 = \frac{32}{3} \, \max_{\{ \zeta_n \}} \, \sum_{n=1}^{\infty}
\frac{\big[ \int_{0}^{\infty} \, \zeta_{n}(r) \, \exp(-r) \, r^4 \, dr \big]^4}
{1-\int_{0}^{\infty} \zeta_n(r) \, [(2 \, \zeta_n(r) + 4 \, \zeta_n'(r) + r \, \zeta_n''(r)] \, r^3 \, dr} \quad .
\end{eqnarray}
\end{widetext}
The reduced occupation numbers of the NOs constructed from $\bar\phi_{nx}(\vec r)$, $\bar\phi_{ny}(\vec r)$, and $\bar\phi_{nz}(\vec r)$ according to Eq.~(\ref{a15}) equal, $\bar \nu_ {nx} = \bar \nu_ {ny} = \tilde \nu_ {n}$, and $\bar \nu_ {nz} = 4 \, \tilde\nu_ {n}$, respectively, where
$\tilde \nu_ {n} = \frac{8}{9} \, \Big( \frac{[\int_{0}^{\infty} \zeta_{n}(r) \, \exp(-r) \, r^4 \, dr ]^2}
{1-\int_{0}^{\infty} \zeta_n(r) \, [(2 \, \zeta_n(r) + 4 \, \zeta_n'(r) + r \, \zeta_n''(r)] \, r^3 \, dr} \Big)^2$.

A practical approach to the computation of $C_6$ involves expanding $\{ \zeta_n(r) \}$ into a finite set $\{ f_k(r) \}$ of $N$ suitable basis functions. A convenient choice for $\{ f_k(r) \}$ is that related to the associated Laguerre polynomials, i.e. \cite{15}
\begin{eqnarray}\label{a28}
f_k(r) &=& 2^{-3/2} \, \frac{[(k-1)!]^{1/2}}{[(k+3)!]^{3/2}} \, \exp(-r) \times  \nonumber \\
&\times& \frac{d^4}{d r^4}  \Big( \hspace{-3pt} \exp(2r) \,
\frac{d^{k+3} \, [r^{k+3} \, \exp(-2r)]}{d r^{k+3}} \Big)  \quad  .
\end{eqnarray}
Thanks to the identity $\int_{0}^{\infty} f_{k}(r) \, \exp(-r) \, r^4 \, dr = \frac{\sqrt{3}}{2} \, \delta_{k1}$, setting
$\zeta_n(r) = \sum_{k=1}^N U_{nk} \,  f_{k}(r)$ for $n =1, \, 2, \, \dots , \, N$, where $\{ U_{nk} \}$ are the elements of an orthogonal matrix $\bm{U}$, produces  the approximation
\begin{widetext}
\begin{eqnarray}\label{a29}
 C_6 \approx C_6^{(N)} = 30
 \, \max_{\bm{U}} \, \sum_{n=1}^{N} \frac{U_{n1}^4} {\sum_{k_1=1}^{N} \sum_{k_2=1}^{N} U_{nk_1} \, U_{nk_2} \, (4 \,k_<+1) \,
\sqrt\frac{k_< \, (k_<+1) \, (k_<+2) \, (k_<+3)}{k_> \, (k_>+1) \, (k_>+2) \, (k_>+3)}} \quad ,
\end{eqnarray}
\end{widetext}
where $k_< = \min(k_1,k_2)$ and $k_> = \max(k_1,k_2)$.  The most straightforward (albeit certainly not the most efficient) numerical approach to finding $\bm{U}$ involves equating it to a product of $2 \times 2$ Jacobi rotations whose angles are determined by repeated maximizations of the r.h.s. of
Eq. (\ref{a29}).  This iterative process is terminated when the estimated error in the computed $C_6^{(N)}$ falls below a prescribed threshold.  The resulting $\bm{U}$ is employed in the construction of the approximates for $N$ radial components $\{ \zeta_n(r) \}$ of the half-space orbitals and the corresponding values of $\{ \tilde \nu_n \}$, the latter being given by
\begin{widetext}
\begin{eqnarray}\label{a30}
 \tilde \nu_n \approx \tilde \nu_n^{(N)}
= \frac{25}{2} \, \frac{U_{n1}^4} {\Big[ \sum_{k_1=1}^{N} \sum_{k_2=1}^{N} U_{nk_1} \, U_{nk_2} \, (4 \,k_<+1) \,
\sqrt\frac{k_< \, (k_<+1) \, (k_<+2) \, (k_<+3)}{k_> \, (k_>+1) \, (k_>+2) \, (k_>+3)} \, \Big]^2} \quad .
\end{eqnarray}
\end{widetext}
Table 1. The properties at $R \to \infty$ of the ten half-space orbitals  with the largest reduced occupation numbers $^{a)}$.


\begin{tabular}{r l r r l}
\hline \hline \noalign{\smallskip}
$n$ & \multicolumn{1}{c}{$\tilde\nu_n \,^{b)}$}	& \multicolumn{1}{c}{$t_n \,^{c)}$} & \multicolumn{1}{c}{$v_n \,^{d)}$}
& \multicolumn{1}{c}{$\Delta C_{6,n} \,^{e)}$}  \\
\hline \noalign{\smallskip}
 1	&	6.164 536$\cdot 10^{-1}$	        &	 0.366 020	        &	0.426 876	&	6.497 086                           \\
 2	&	9.835 517$\cdot 10^{-5}$	        &	 0.797 570	        &	0.477 839	&	1.934 996$\cdot 10^{-3}$	 \\
 3	&	1.603 983$\cdot 10^{-7}$	        &	 1.442 727	        &	0.551 436	&	5.355 857$\cdot 10^{-6}$	 \\
 4	&	8.051 477$\cdot 10^{-10}$	&	 2.347 228	        &	0.634 729	&	4.275 332$\cdot 10^{-8}$	 \\
 5	&	8.023 493$\cdot 10^{-12}$	&	 3.583 218	        &	0.727 563	&	6.461 778$\cdot 10^{-10}$	 \\
 6	&	1.282 445$\cdot 10^{-13}$	&	 5.242 253	        &	0.830 473	&	1.511 780$\cdot 10^{-11}$	 \\
 7	&	2.910 896$\cdot 10^{-15}$	&	 7.437 599	        &	0.944 096	&	4.885 767$\cdot 10^{-13}$	 \\
 8	&	8.687 078$\cdot 10^{-17}$	&	10.307 968	&	1.069 137	&	2.030 448$\cdot 10^{-14}$	 \\
 9	&	3.234 794$\cdot 10^{-18}$	&	14.022 087	&	1.206 360	&	1.033 767$\cdot 10^{-15}$	 \\
10	&	1.447 666$\cdot 10^{-19}$	&	18.784 099	&	1.356 590	&	6.228 730$\cdot 10^{-17}$	 \\

\noalign{\smallskip} \hline \hline

\end{tabular}

\noindent $^{a)}$ The approximate values computed at $N=200$.

\noindent $^{b)}$ $\tilde\nu_n = \bar\nu_{nx} = \bar\nu_{ny} = \frac{1}{4} \, \bar\nu_{nz}$.

\noindent $^{c)}$ $t_n = \langle \bar\phi_{nx}(\vec r) | - \frac{1}{2} \, \hat\nabla^2 | \bar\phi_{nx}(\vec r) \rangle
                    = \langle \bar\phi_{ny}(\vec r) | - \frac{1}{2} \, \hat\nabla^2 | \bar\phi_{ny}(\vec r) \rangle$

                    $\hspace{7pt} = \langle \bar\phi_{nz}(\vec r) | - \frac{1}{2} \, \hat\nabla^2 | \bar\phi_{nz}(\vec r) \rangle$.

\noindent $^{d)}$ $v_n = \langle \bar\phi_{nx}(\vec r) |\hat r^{-1} | \bar\phi_{nx}(\vec r) \rangle
                    = \langle \bar\phi_{ny}(\vec r) | \hat r^{-1} | \bar\phi_{ny}(\vec r) \rangle
                    = \langle \bar\phi_{nz}(\vec r) |\hat r^{-1} | \bar\phi_{nz}(\vec r) \rangle$.

\noindent $^{e)}$ The collective contribution $\Delta C_{6,n} = 12 \, \tilde\nu_n [1+2 \, (t_n-v_n)]$ of the half-space orbitals $\bar\phi_{nx}$,
$\bar\phi_{ny}(\vec r)$, and $\bar\phi_{nz}(\vec r)$ to $C_6$.
\\
\\
For $N=1$, Eqs.(\ref{a29}) and (\ref{a30}) trivially yield the approximates $C_6^{(1)}=6$ and $\tilde \nu_1^{(1)}=\frac{1}{2}$, whereas for $N=2$ one obtains $C_6^{(2)}=\frac{363}{56} \approx 6.482 \, 143$ and $\{ \tilde \nu_1^{(2)},  \tilde \nu_2^{(2)} \}=\{ \frac{3869 + 63 \, \sqrt{3769}}{12544}, \frac{3869 - 63 \, \sqrt{3769}}{12544} \} \approx \{ 0.616 \, 766, 1.030 \, 405 \hspace{-1pt} \cdot \hspace{-1pt} 10^{-4} \} $, the latter being close to the previously published (rather inaccurate) estimates of $\{ \frac{1}{2} \hspace{-1pt} \cdot \hspace{-1pt} 1.110 \, 371^2, \frac{1}{2} \hspace{-1pt} \cdot \hspace{-1pt} 0.014 \, 029^2 \} = \{ 0.616 \, 462, 0.984 \, 064 \hspace{-1pt} \cdot \hspace{-1pt} 10^{-4}\}$ \cite{15}.  The convergence of $C_6^{(N)}$ with $N$ is quite rapid as attested by the computed values of
$C_6^{(20)}   \approx 6.499 \, 026 \, 705 \, 405 \, 836$,
$C_6^{(50)}   \approx 6.499 \, 026 \, 705 \, 405 \, 839 \, 313 \, 125$, and
$C_6^{(100)} \approx 6.499 \, 026 \, 705 \, 405 \, 839 \, 313 \, 128 \, 194 \, 56$
that, when juxtaposed against the best available estimate of
$C_6 \approx  6.499 \, 026 \, 705 \, 405 \, 839 \, 313 \, 128 \, 194 \, 604 \, 1$ \cite{18},
are found to match, respectively, 15, 21, and 27 of its digits.  At least 30 digits of
$C_6^{(200)} \approx 6.499 \, 026 \, 705 \, 405 \, 839 \, 313 \, 128 \, 194 \, 604 \, 13$ are correct.

In contrast, the computed approximate properties of the half-space orbitals converge rather slowly with $N$.  For example, calculations with $N=200$ yield sufficiently accurate values of these properties for only the first 25 half-space orbitals with the largest reduced occupation numbers.  This behavior stems from the rapid decay of $\tilde \nu_n$ with $n$ and thus also that of the collective contributions of $\bar\phi_{nx}$, $\bar\phi_{ny}(\vec r)$, and $\bar\phi_{nz}(\vec r)$ to $C_6$, which means that the approximates $C_6^{(N)}$ are quite insensitive to the errors in the properties of the half-space orbitals with large ordinal numbers.  In other words, highly accurate estimates of $C_6$ are obtained from rather inaccurate half-space orbitals and their reduced occupation numbers.  This phenomenon makes the maximization of the r.h.s. of Eq.~(\ref{a29}) numerically ill-conditioned for large $N$, necessitating the employment of arbitrary-precision arithmetic (e.g.~1200 digits for $N=200$) incorporated in appropriate software \cite{48}.  A selection of the data obtained from such calculations is presented in Table 1.

\section{PROPERTIES OF THE HALF-SPACE ORBITALS AT $R \to \infty$}
\begin{figure}[htb]
 \centering
  \includegraphics[width=7cm, clip=true, trim = 0.0cm 1.9cm 1.2cm 0.0cm]{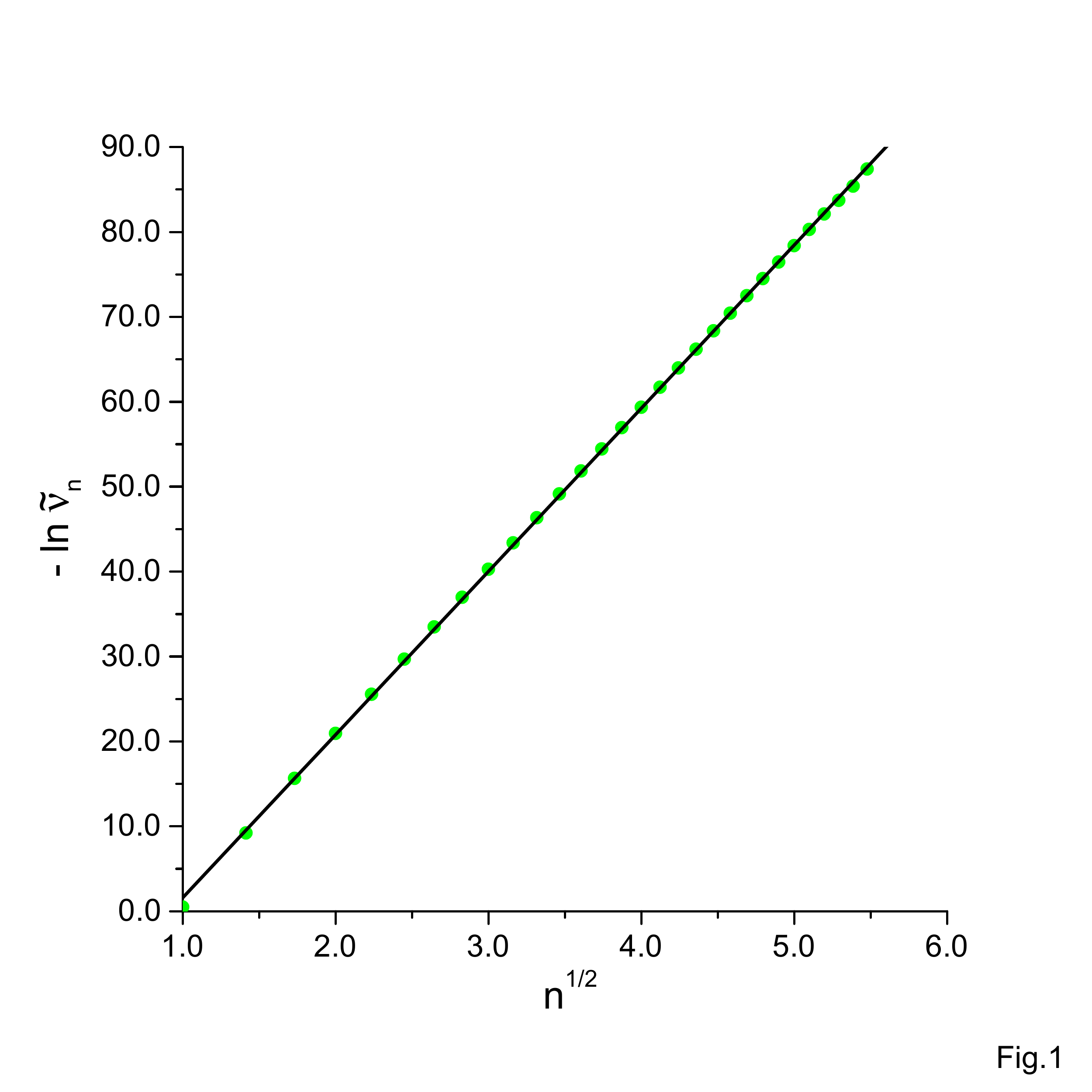}
\caption{$- \ln \tilde\nu_n$ vs. $n^{1/2}$.  The data computed at $N=200$; the line is provided for eye guidance only.}  \label{fig:Fig 1}
\end{figure}
The occupation numbers of the NOs pertaining to a singlet state of a Coulombic system are known to obey the asymptotic power law \cite{32,33}
\begin{eqnarray}\label{a31}
\lim_{\mathfrak{n} \to \infty} \, \mathfrak{n}^{8/3} \, \nu_{\mathfrak{n}} = \Big( \frac{2^{1/2}}{3 \, \pi^{5/4} } \, \int \, [\rho_2(\vec r,\vec r)]^{3/8} \; d\vec r \, \Big)^{\hspace{-2pt}8/3} \ ,
\end{eqnarray}
where $\rho_2(\vec r,\vec r)$ is the respective on-top two-electron density.  In the present case of two ground-state hydrogen atoms with the internuclear separation $R$, the large-$R$ asymptotics of $\big( \int \, [\rho_2(\vec r,\vec r)]^{3/8} \; d\vec r \, \big)^{8/3}$ is given by
$\frac{4096}{729} \,  (6 \, \pi)^{2/3} \, (1 + \frac{R}{2})^{16/3} \, \exp(-2 \,R) = \frac{4096}{729} \,  (6 \, \pi)^{2/3} \, (1 + \frac{R}{2})^{10/3} \,  \mathfrak{P}^2$ \cite{44}.  When employed in conjunction with Eq. (\ref{a31}), this asymptotics implies $\lim_{n \to \infty} \, n^{8/3} \, \tilde\nu_n$ being of the order of $R^{34/3} \, \exp(-2 \,R)$, i.e. vanishing at the limit of $R \to \infty$.  In other words, the reduced occupation numbers $\{ \tilde\nu_n \}$ decay faster than $n^{-8/3}$ at this limit.  As the exact nature of this decay cannot be ascertained within the formalism employed in the derivation of Eq. (\ref{a31}), it has to be examined with numerical methods.

\begin{figure}[htb]
 \centering
  \includegraphics[width=7cm, clip=true, trim = 0.0cm 1.9cm 1.2cm 0.0cm]{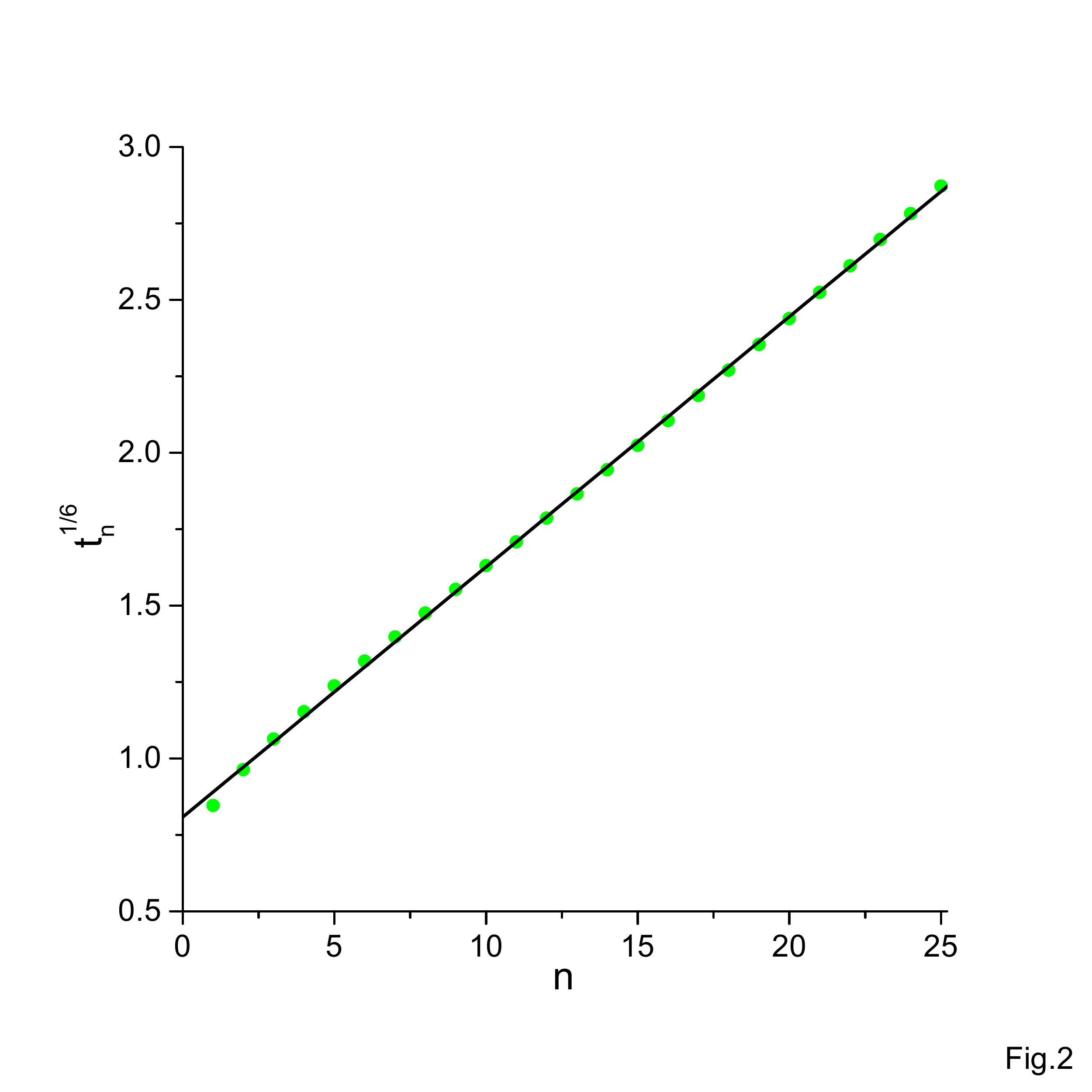}
\caption{ $t_n^{1/6}$ vs. $n$.  The data computed at $N=200$; the line is provided for eye guidance only.}  \label{fig:Fig 2}
\end{figure}
\begin{figure}[htb]
 \centering
  \includegraphics[width=7cm, clip=true, trim = 0.0cm 1.9cm 1.2cm 0.0cm]{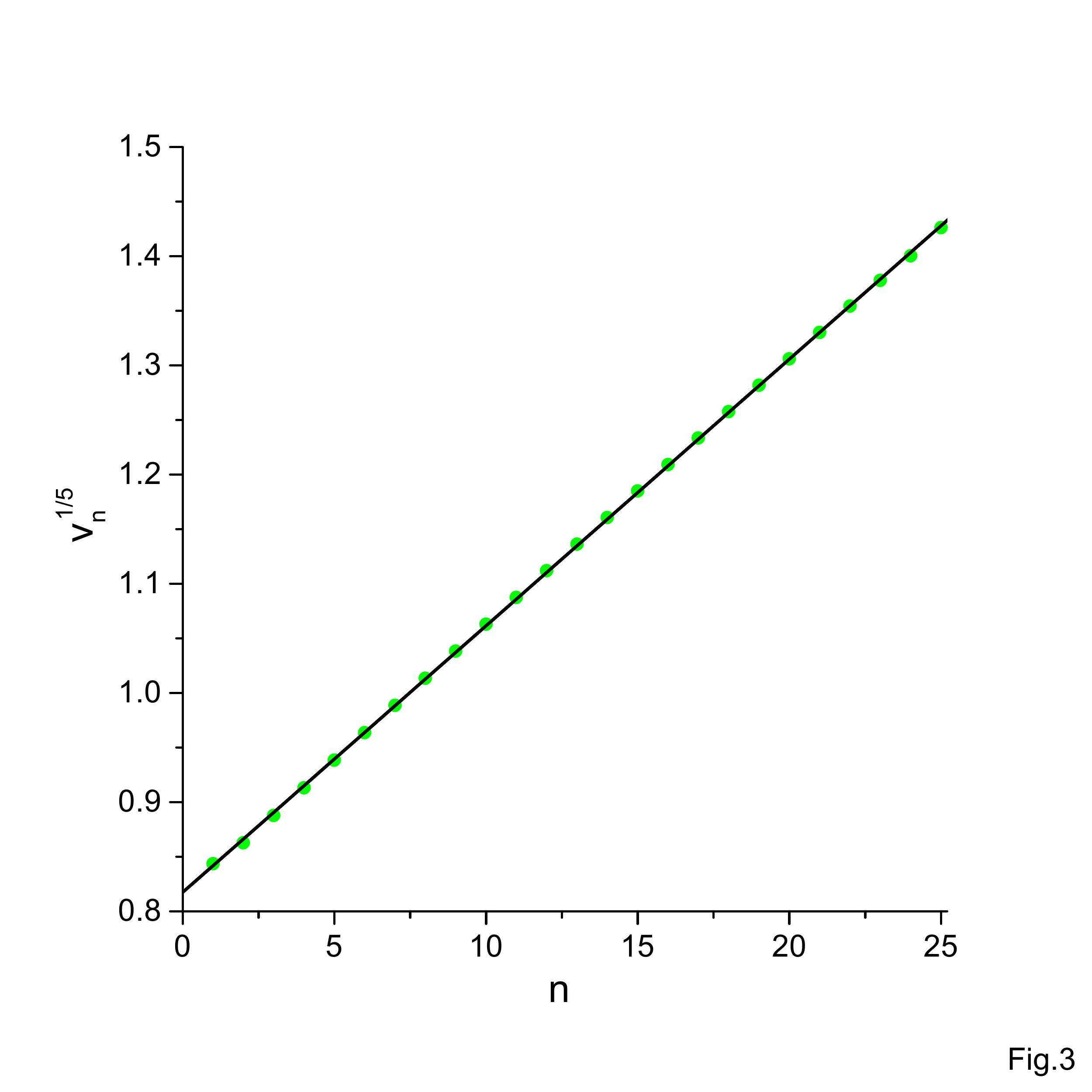}
\caption{ $v_n^{1/5}$ vs. $n$.  The data computed at $N=200$; the line is provided for eye guidance only.
 }  \label{fig:Fig 3}
\end{figure}
Inspection of Fig.~1, in which accurate values of the reduced occupation numbers computed at $N=200$ are displayed, reveals the subexponential decay of
$\{ \tilde\nu_n \}$ with $n$ that follows the approximate formula $\tilde\nu_n \approx 4.504 \cdot 10^{7} \, \exp(-19.216 \, n^{1/2})$.  On the other hand, the expectation values
$t_n = \langle \bar\phi_{nx}(\vec r) | - \frac{1}{2} \, \hat\nabla^2 | \bar\phi_{nx}(\vec r) \rangle = \langle \bar\phi_{ny}(\vec r) | - \frac{1}{2} \, \hat\nabla^2 | \bar\phi_{ny}(\vec r) \rangle = \langle \bar\phi_{nz}(\vec r) | - \frac{1}{2} \, \hat\nabla^2 | \bar\phi_{nz}(\vec r) \rangle$ and
$v_n = \langle \bar\phi_{nx}(\vec r) |\hat r^{-1} | \bar\phi_{nx}(\vec r) \rangle = \langle \bar\phi_{ny}(\vec r) | \hat r^{-1} | \bar\phi_{ny}(\vec r) \rangle
= \langle \bar\phi_{nz}(\vec r) |\hat r^{-1} | \bar\phi_{nz}(\vec r) \rangle$ appear to scale, respectively, like $n^6$ and $n^5$ for large $n$ (Figs.~2 and 3).  This growth of the expectation values with $n$ is much steeper than that observed in the case of the ground-state helium atom \cite{32}.
\begin{figure}[htb]
 \centering
  \includegraphics[width=7cm, clip=true, trim = 0.0cm 1.5cm 1.2cm 0.0cm]{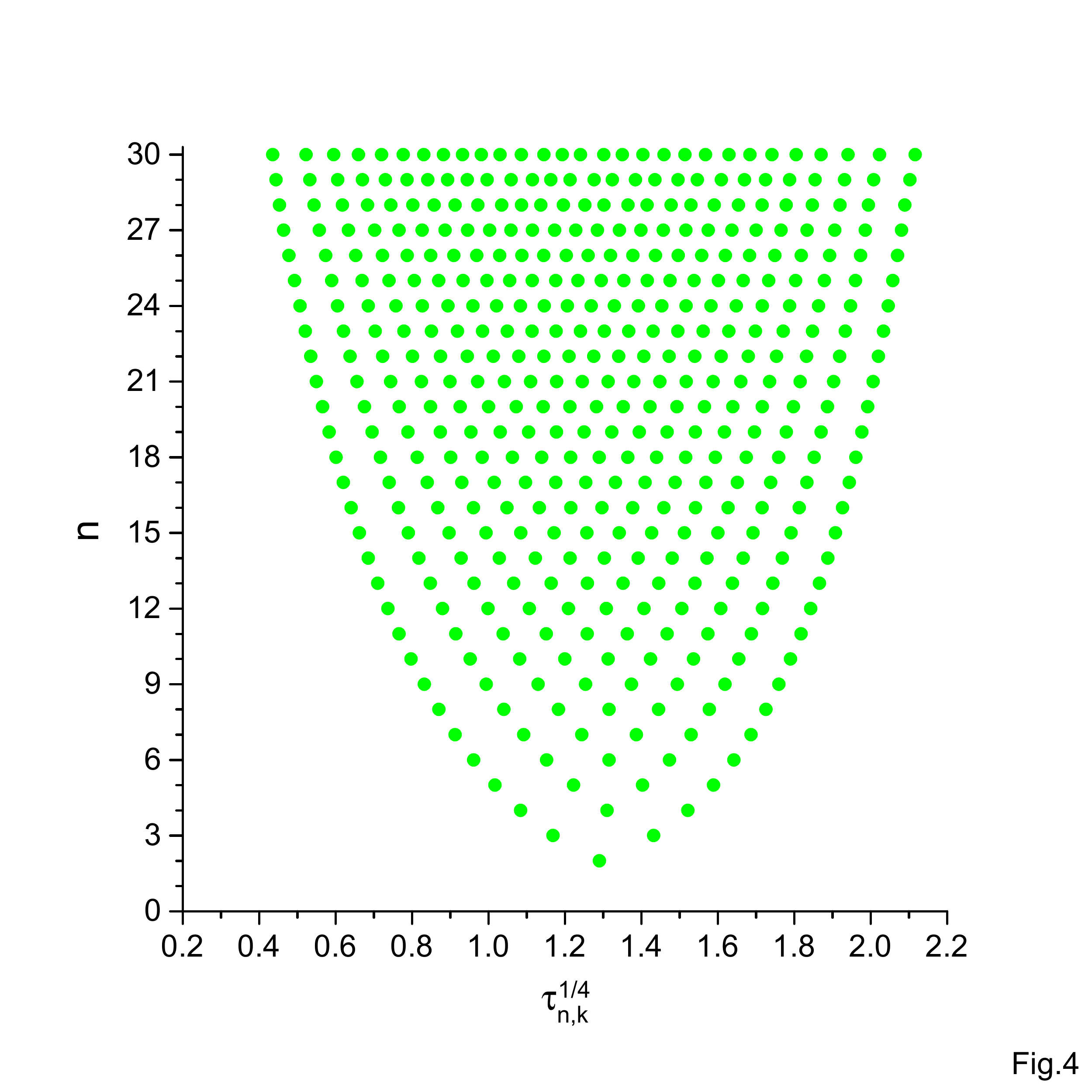}
\caption{ The nodes $\{ \tau_{n,k} \}$ of the radial components $\{ \zeta_n(r) \}$ .  The data computed at $N=200$ for $2 \le n \le 30$. }  \label{fig:Fig 4}
\end{figure}
\begin{figure}[htb]
 \centering
  \includegraphics[width=7cm, clip=true, trim = 0.0cm 1.9cm 1.2cm 0.0cm]{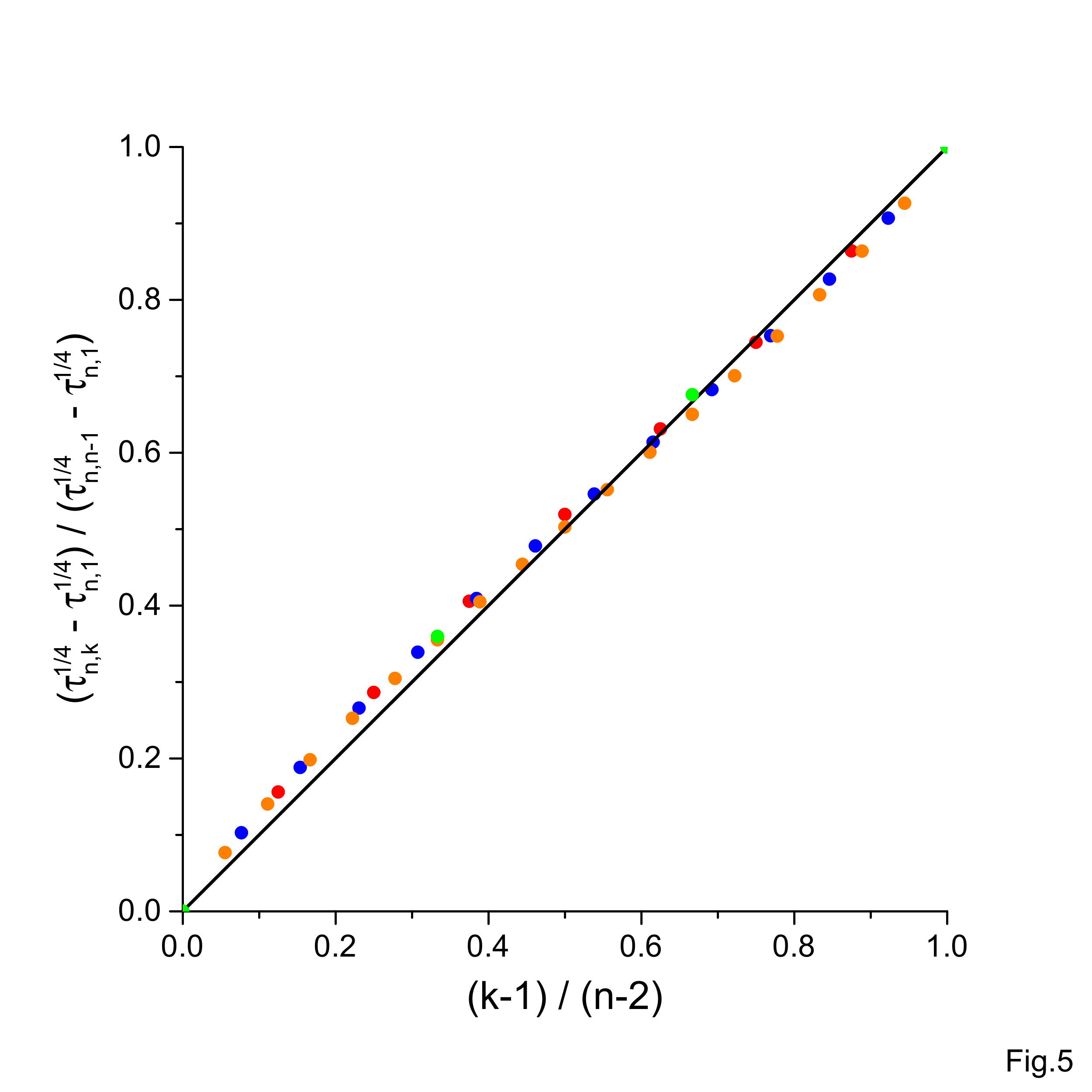}
\caption{ $\frac{\tau^{1/4}_{n,k}-\tau^{1/4}_{n,1}}{\tau^{1/4}_{n,n-1}-\tau^{1/4}_{n,1}}$ vs. $\frac{k-1}{n-2}$ for $n=5$ (green), 10 (red), 15 (blue), and 20 (orange) .  The data computed at $N=200$; the line is provided for eye guidance only.}  \label{fig:Fig 5}
\end{figure}
\begin{figure}[htb]
 \centering
  \includegraphics[width=7cm, clip=true, trim = 0.0cm 1.7cm 1.2cm 0.0cm]{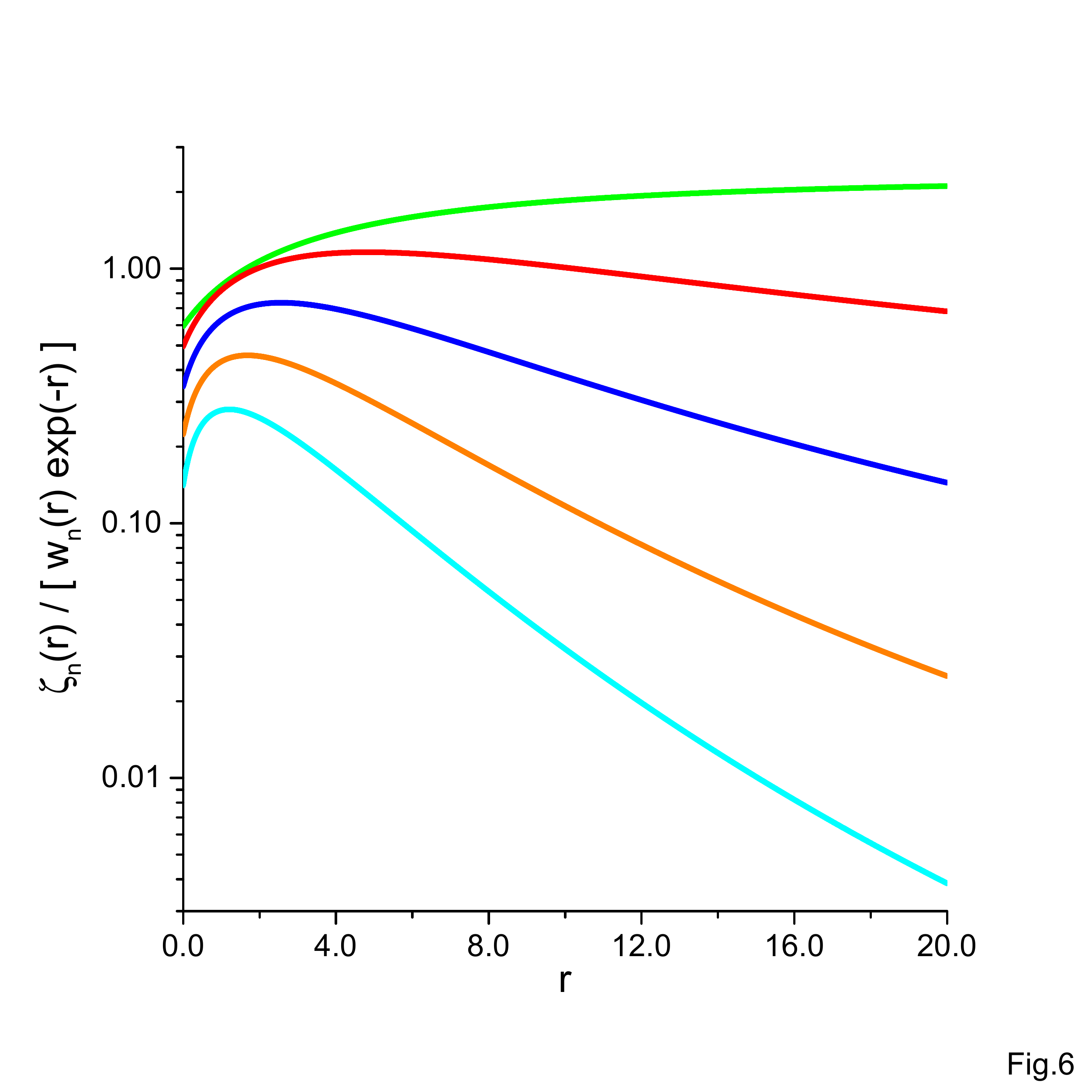}
\caption{$\frac{\zeta_n(r)}{w_n(r) \exp(-r)}$ vs. $r$ for $n=1$ (green), 2
(red), 3 (blue), 4 (orange), and 5 (cyan).  The data computed at $N=200$.}
 \label{fig:Fig 6}
\end{figure}

The nodes $\{ \tau_{n,k} \}$ of the radial components $\{ \zeta_n(r) \}$ of the half-space orbitals, defined as
$\forall_{n} \, \forall_{1 \ge k \ge n-1} \, \zeta_n(\tau_{n,k}) = 0$, exhibit regularities that go beyond the simple interleaving (Fig.~4).  In particular, the (nonlinear) relationship between the ratios $\frac{k-1}{n-2}$ and $\frac{\tau^{1/4}_{n,k}-\tau^{1/4}_{n,1}}{\tau^{1/4}_{n,n-1}-\tau^{1/4}_{n,1}}$ appears to be approximately $n$-independent (Fig.~5).

At the first glance, the main features of $\zeta_n(r)$ are expected to be embodied in the function $w_n(r) \, \exp(-r)$, where
$w_n(r) = \Pi_{k=1}^{n-1} \, (r-\tau_{n,k})$, that reproduces all its nodes and accounts for its expected \cite{49} large-$r$ asymptotics of $\exp(-r)$.  However, the plots of the ratios $\frac{\zeta_n(r)}{w_n(r) \exp(-r)}$ (Fig.~6) reveal for $n>1$ the presence of residual $n$-dependent exponential decay at
$r \to \infty$. The origin of this unexpected behavior remains unknown at present.

\section{DISCUSSION AND CONCLUSIONS}

The nonrelativistic energy of the ground-state H$_2$ molecule is given within the clamped-nuclei approximation by a functional of the pertinent one-electron reduced density matrix (the 1-matrix) that becomes asymptotically free of the phase dilemma at the limit of infinite internuclear separation $R$.  The large-$R$ behavior of the respective natural orbitals (NOs) is characterized by (asymptotically perfect) pairing of their occupation numbers, each pair involving natural amplitudes with opposite signs that correspond to the NOs with opposite parities.  These NOs arise from symmetric and antisymmetric combinations of ''half-space'' orbitals that are asymptotically unique for each pair.  The natural amplitudes pertaining to the symmetric combinations exhibit asymptotically fixed sign pattern of being positive-valued for the NO with the occupation number close to one-half and negatively-valued otherwise.  The large-$R$ asymptotics of the occupation numbers of the weakly occupied NOs have the leading terms proportional to $R^{-n}$, where $n=6$ for the NOs composed of the $p$-type half-space orbitals and $n > 6$ for the others.  The multiplicative factors in these terms (i.e. the reduced occupation numbers) are given by explicit expressions involving the half-space orbitals.

Minimization of the 1-matrix functional for the electronic energy is asymptotically equivalent to maximization of the functional for the $C_6$ dispersion coefficient with respect to an orthonormal set of the radial components of the half-space orbitals.  Introduction of a finite basis set facilitates computations of $C_6$ with standard numerical methods, the accuracy corresponding to at least 30 correct digits being readily attainable with 200 basis functions.  The resulting benchmark-quality approximates for the radial components of several half-space orbitals and the corresponding reduced occupation numbers, available for the first time thanks to the development of the present formalism, turn out to exhibit several unexpected properties that elude analytical explanation at present and thus warrant further investigation.

The capability of the present formalism to afford the $C_6$ estimates of arbitrarily high accuracy (easily matching or exceeding that obtained previously with a variety of methods \cite{11,12,13,14,15,16,17,18}) demonstrates that essentially exact numerical values of various quantities can be computed within the one-electron reduced density matrix functional theory.  One example of such quantities is the occupation numbers $\nu_{\mathfrak n}$ of the weakly occupied natural orbitals whose aforedescribed scaling with $R$ leads to the conclusion that the von Neumann entropy
$S = -\sum_{\mathfrak{n}} \nu_{\mathfrak n} \ln \nu_{\mathfrak n}$ (as well as its variant exhibiting the particle-hole symmetry \cite{50}) has the leading large-$R$ asymptotics given by a sum of a constant and a term proportional to $R^{-6} \, \ln R$.  Similarly, one concludes that the so-called cumulant energy $E_{cum}$ \cite{50} scales like $R^{-6}$ for large values of $R$ \cite{51}.  These observations unequivocally prove that one of the formulations of the Collins conjecture \cite{52} i.e. that of a linear relationship between $S$ and $E_{cum}$ \cite{50} is not valid within the large-$R$ regime.

Several extensions of the approach presented in this paper are possible.  Among them, the derivation of expressions for higher dispersion coefficients
$\{C_{2n} \}, n=4,5,\dots$ for the H$_2$ molecule (via a systematic perturbational approach to the solution of the variational equations of the energy functional) and the computation of $C_6$ for other two-electron systems, such as the $H_3^+$ cation are worth mentioning here.

\begin{acknowledgments}
The research described in this publication has been funded by the National Science Center (Poland) under grant 2018/31/B/ST4/00295, the German Research Foundation under grant SCHI 1476/1-1, and the Munich Center for Quantum Science and Technology. It is also a part of the Munich Quantum Valley program, which is supported by the Bavarian state government with funds from the Hightech Agenda Bayern Plus.
\end{acknowledgments}


\begin{thebibliography}{99}
\bibitem{1} Note that the term ''van der Waals interaction'' is not used consistently in the literature.  On one hand, per IUPAC recommendations \cite{2}, it is defined as "The attractive or repulsive forces between molecular entities (or between groups within the same molecular entity) other than those due to bond formation or to the electrostatic interaction of ions or of ionic groups with one another or with neutral molecules. The term includes: dipole–dipole, dipole-induced dipole and London (instantaneous induced dipole-induced dipole) forces. The term is sometimes used loosely for the totality of nonspecific attractive or repulsive intermolecular forces.", which is confusing as the dipole-dipole interactions \textit{are} electrostatic interactions.  On the other hand, see e.g. the titles of the seminal papers on the dispersion interactions \cite{3,4,9}.
\bibitem{2} P. Muller, Pure Appl. Chem. \textbf{66}, 1077 (1994).
\bibitem{3} E. H. Lieb and W. E. Thirring, Phys. Rev. A \textbf{34}, 40 (1986).
\bibitem{4} H. B. G. Casimir and D. Polder, Phys. Rev. \textbf{73}, 360 (1948).
\bibitem{5} C. Mavroyannis and M. J. Stephen, Mol. Phys. \textbf{5}, 629 (1962).
\bibitem{6} F. London, Z. Phys. Chem., Abt. B \textbf{11}, 222 (1930).
\bibitem{7} F. London, Z. Phys. \textbf{63}, 245 (1930).
\bibitem{8} D. P. Kooi and P. Gori-Giorgi, J. Phys. Chem. Lett. \textbf{10}, 1537 (2019).
\bibitem{9} R. Eisenchitz and F. London, Z. Phys. \textbf{60}, 491 (1930).
\bibitem{10} J. C. Slater and J. G. Kirkwood, Phys. Rev. \textbf{37}, 682 (1931).
\bibitem{11} T. C. Choy, Phys. Rev. A \textbf{62}, 012506 (2000).
\bibitem{12} E. Canc\`{e}s and L. R. Scott, SIAM J. Math. Anal. \textbf{50}, 381 (2018).
\bibitem{13} L. Pauling and J. Y. Beach, Phys. Rev. \textbf{47}, 686 (1935).
\bibitem{14} Y. M. Chan and A. Dalgarno, Mol. Phys. \textbf{9}, 349 (1965).
\bibitem{15} J. O. Hirschfelder and P.-O. L\"owdin, Mol. Phys. \textbf{2}, 229 (1959); \textbf{9}, 491 (1965) (E).
\bibitem{16} R. J. Bell, Proc. Philos. Soc. London \textbf{87}, 594 (1966).
\bibitem{17} A. J. Thakkar, J. Chem. Phys. \textbf{89}, 2092 (1988).
\bibitem{18} M. Masili and R. J. Gentil, Phys. Rev. A \textbf{78}, 034701 (2008).
\bibitem{19} T. L. Gilbert, Phys. Rev. B \textbf{12}, 2111 (1975).
\bibitem{20} R. A. Donnelly and R. G. Parr, J. Chem. Phys. \textbf{69}, 4431 (1978).
\bibitem{21} S. M. Valone, J. Chem. Phys. \textbf{73}, 1344 (1980).
\bibitem{22} M. Levy, Proc. Natl. Acad. Sci. \textbf{76}, 6062 (1979).
\bibitem{23} For a recent review see: K. Pernal and K. J. H. Giesbertz, Top. Curr. Chem. \textbf{368}, 125 (2016).
\bibitem{24} J. Liebert, F. Castillo, J.-P. Labb\'{e}, and C. Schilling, J. Chem. Theory Comput. \textbf{18}, 124 (2022).
\bibitem{25} C. L. Benavides-Riveros, J. Wolff, M. A. L. Marques, and C. Schilling, Phys. Rev. Lett. \textbf{124}, 180603 (2022).
\bibitem{26} J. Liebert and C. Schilling, "Functional Theory for Excitations in Boson Systems", arXiv:2204.12715 (2022).
\bibitem{27} J. Cioslowski, Z. \'{E}. Mih\'{a}lka, and \'{A}. Szabados, J. Chem. Theory Comput. \textbf{15}, 4862 (2019).
\bibitem{28} J. Cioslowski, J. Chem. Theory Comput. \textbf{16}, 1578 (2020).
\bibitem{29} C. Schilling, J. Chem. Phys. \textbf{149}, 231102 (2018).
\bibitem{30} K. Pernal and J. Cioslowski, J. Chem. Phys. \textbf{120}, 5987 (2004).
\bibitem{31} J. Cioslowski and F. Pr\c{a}tnicki, J. Chem. Phys. \textbf{151}, 184107 (2019).
\bibitem{32} J. Cioslowski and K. Strasburger, J. Chem. Theory Comput. \textbf{17}, 6918 (2021).
\bibitem{33} A. V. Sobolev, Funct. Anal. Appl. \textbf{55}, 113 (2021).
\bibitem{34} P.-O. L\"owdin and H. Shull, Phys. Rev. \textbf{101}, 1730 (1956).
\bibitem{35} J. Cioslowski, J. Chem. Phys. \textbf{148}, 134120 (2018).
\bibitem{36} S. Goedecker and C. J. Umrigar, in \textit{Many-Electron Densities and Reduced Density Matrices}, edited by J. Cioslowski (Kiuwer Academic, Dordrecht/New York, 2000), Chap. 8, pp. 165-181.
\bibitem{37} J. Cioslowski and M. Buchowiecki, J. Chem. Phys. \textbf{125}, 064105 (2006).
\bibitem{38} J. Cioslowski and K. Pernal, J. Chem. Phys. \textbf{113}, 8434 (2000).
\bibitem{39} J. Cioslowski and F. Pr\c{a}tnicki, J. Chem. Phys. \textbf{153}, 224106 (2020).
\bibitem{40} J. Cioslowski and K. Pernal, Chem. Phys. Lett. \textbf{430}, 188 (2006).
\bibitem{41} D. D. Konowalow, W. H. Barker, and R. Mandel, J. Chem. Phys. \textbf{49}, 5137 (1968).
\bibitem{42} X. W. Sheng, \L . M. Mentel, O. V. Gritsenko, and E. J. Baerends, J. Chem. Phys. \textbf{138}, 164105 (2013).
\bibitem{43} J. Cioslowski, F. Pratnicki, and K. Strasburger, J. Chem. Phys. \textbf{156}, 034108 (2022).
\bibitem{44} For example, $\mathfrak{P}$ computed from the FCI wavefunction involving the one-electron basis set
$\{ \chi(\vec r),\chi(\vec r^{\;\bullet}) \}$, where $\chi(\vec r)$ is the 1s orbital of the hydrogen atom (with either the original or optimized exponent) centered at  $\frac{1}{2} \, R \, \vec e_z$, has the large-$R$ asymptotics of $(1+\frac{R}{2}) \, \exp(-R)$ that, contrary to simplistic expectations, is equal to
$2 \, \langle \chi(\vec r)|\theta(-\vec e_z \cdot \vec r)|\chi(\vec r) \rangle$ rather than $\langle \chi(\vec r)|\chi(\vec r^{\;\bullet}) \rangle$.
\bibitem{45} R. J. Buehler and J. O. Hirschfelder, Phys. Rev. \textbf{83}, 628 (1951).
\bibitem{46} This assertion follows from a straightforward variational argument.  Since the 1s orbital of the hydrogen atom is the lowest-energy eigenfunction of the operator $ - \frac{1}{2} \, \hat \nabla_1^2 - (|\vec r|^{-1} + \frac{1}{2} \, R^{-1})  \, \hat 1$ whose finite-valued expectation values are
$\{  h_{\mathfrak{n}\mathfrak{n}} \}$, $\forall_{\mathfrak{n} \ne 0} \; h_{\mathfrak{n}\mathfrak{n}} > h_{00}$ at the limit of $R \to \infty$.  Moreover, one should note that $I_{00} = 2 \, \big( \langle \bar\phi_0(\vec r)|\hat z| \bar\phi_0(\vec r) \rangle \big)^2$, which measures the deviation of $\bar\phi_0(\vec r)$ from sphericity, decays faster than $R^{-3}$ as $R \to \infty$ \cite{15}.
\bibitem{47} O. Perron, Math. Ann. \textbf{64}, 248 (1907).
\newline G. Frobenius, Sitzungsber. Kgl. Preuss. Akad. Wiss., 471 (1908), 514 (1909), 456 (1912).
\bibitem{48} Mathematica, Version 12.2.0.0, Wolfram Research, Inc., Champaign, IL, 2020.
\bibitem{49} J. Katriel and E. R. Davidson, Proc. Natl. Acad. Sci. U. S. A. \textbf{77}, 4403 (1980).
\bibitem{50} J. Wang and E. J. Baerends, Phys. Rev. Lett. \textbf{128}, 013001(2022) and the references cited therein.
\bibitem{51} M. Via-Nadal, M. Rodríguez-Mayorga, and E. Matito, Phys. Rev. A \textbf{96}, 050501 (2017).
\newline M. Via-Nadal, M. Rodríguez-Mayorga, E. Ramos-Cordoba, and E. Matito, J. Phys. Chem. Lett. \textbf{10}, 4032 (2019).
\newline  O. Werba, A. Raeber, K. Head-Marsden, and D. A. Mazziotti, Phys. Chem. Chem. Phys. \textbf{21}, 23900 (2019).
\bibitem{52} D. M. Collins, Z. Naturforsch. \textbf{48A}, 68 (1993).
\end{thebibliography}
\end{document}